\documentclass[twocolumn,twoside]{IEEEtran}

\ifCLASSOPTIONcompsoc
\usepackage[caption=false,font=normalsize,labelfont=sf,textfont=sf]{subfig}
\else
\usepackage[caption=false,font=normalsize]{subfig}
\fi

\usepackage[dvipsnames]{xcolor}
\usepackage{balance}
\usepackage{cite}
\usepackage{gensymb}
\usepackage{multirow}
\usepackage{graphicx}
\usepackage{epstopdf}
\usepackage{textcomp}
\usepackage{amsmath}
\usepackage{mathtools}
\usepackage{amssymb}
\usepackage{amsfonts}
\usepackage{float}
\usepackage{amsthm}
\usepackage{braket}

\begin{document}

\title{OHL-Assisted All-Optical Regenerative Relaying for Pointing-Impaired M-PAM Inter-Satellite Links}

\author{
Meysam Ghanbari,~Mohammad~Taghi~Dabiri,~Zain~Ali,~Rula~Ammuri,
\\~Mazen~Hasna,~{\it Senior Member,~IEEE},~Iman~Tavakkolnia,~and~Khalid~Qaraqe,~{\it Senior Member,~IEEE}

\thanks{Meysam Ghanbari, Zain Ali, and Khalid~Qaraqe are with the College of Science and Engineering, Hamad Bin Khalifa University, Doha, Qatar (E-mails: megh89467@hbku.edu.qa; zali@hbku.edu.qa; kqaraqe@hbku.edu.qa).}

\thanks{Mohammad Taghi Dabiri and Iman Tavakkolnia are with LiFi R$\&$D centre, Department of Engineering, University of Cambridge, Cambridge, UK (E-mails: md2173@cam.ac.uk; it360@cam.ac.uk).}

\thanks{Rula Ammuri is with Professionals for Smart Technology, Amman, Jordan (E-mail: rammuri@pst.jo).}

\thanks{Mazen Hasna is with the Department of Electrical Engineering, Qatar University, Doha, Qatar (E-mail: hasna@qu.edu.qa).}

\thanks{This work was supported by the Qatar Research Development and Innovation Council (QRDI) under Grant No. NPRP14C-0909-210008 and by research funding from Hamad Bin Khalifa University under the Thematic Research Grant Program Cycle 3. The statements made herein are solely the responsibility of the authors. The content is solely the responsibility of the authors and does not necessarily represent the official views of QRDI.}
}

\maketitle

\begin{abstract}
Rapid inter-satellite traffic growth in LEO constellations demands spectrally efficient, low-latency optical relaying. While amplify and forward (AF) relays are latency-efficient, they propagate noise; conversely, decode and forward (DF) relays suppress noise but incur significant complexity via O/E/O conversion. This paper proposes an all-optical regenerative relay for M-ary pulse amplitude modulation (M-PAM) multi-hop links under pointing errors. A parallel optical hard-limiter (OHL) bank performs symbol-level discrimination, regenerating signal levels directly in the optical domain. A variable gain EDFA is employed to stabilize power and define the threshold-stable region. By incorporating pointing-induced fading, various noise sources including ASE–ASE and signal–ASE beat noise and implementation-dependent decision noise, we derive closed-form per-hop symbol error rate (SER) expressions. These are extended to end-to-end performance using a Markov transition-matrix model for arbitrary modulation order and hop count. Analysis of beamwidth, pointing accuracy, and threshold scaling demonstrates reliable multi-hop operation, avoiding both AF noise accumulation and DF O/E/O processing overhead. Verified by Monte Carlo simulations and numerical integration, this framework provides a design benchmark for low-latency, pointing-aware all-optical regenerative relaying.
\end{abstract}


\IEEEpeerreviewmaketitle

\section{Introduction}

Low-Earth-orbit (LEO) satellite networks are vital for 6G, providing ubiquitous, high-throughput, low-latency connectivity beyond terrestrial reach. Dense constellations support global broadband, resilient backhaul, and real-time routing via inter-satellite links (ISLs), minimizing ground-station reliance. As density increases, radio-frequency ISLs are constrained by spectrum scarcity, interference, antenna-size limitations, and narrow bandwidth. Free-space optical (FSO) ISLs mitigate these challenges by leveraging narrow beams, high carrier frequencies, unlicensed bandwidth, and minimal interference \cite{balakrishnan2025toward,mouhammad2026optimal}. Despite these advantages, the narrow beam directivity that makes FSO ISLs spectrally efficient also makes them sensitive to acquisition, pointing, and tracking errors \cite{ghanbari2026narrowbeams}. In LEO constellations, satellite mobility, platform vibration, attitude-control residuals, and dynamic link geometry can misalign the beam at the receiver aperture, causing power fluctuations, reduced collected energy, and higher symbol-error probability, especially over long inter-satellite distances \cite{dabiri2018channel}. Multi-hop relaying is therefore a natural solution for scalable optical LEO networking, as it divides a long end-to-end path into shorter inter-satellite hops and adds routing flexibility under time-varying constellation geometry \cite{liu2020relay}. Intermediate satellites relax the per-hop link budget, reduce propagation-loss impact, improve route availability, and support long-distance connectivity without a direct source–destination optical line-of-sight link \cite{dabiri2018allopticalaf,erdogan2022secrecy}. However, relay-assisted 6G satellite networks must satisfy strict latency requirements because per-relay delay accumulates along the route. Hence, the relay forwarding mechanism must improve reliability under pointing-impaired optical links without excessive processing, buffering, or conversion delay. Existing relay-assisted satellite communication mainly relies on amplify-and-forward (AF) and decode-and-forward (DF), which trade off implementation simplicity, noise handling, regeneration capability, and latency \cite{choudhary2024isowc}.

DF relaying regenerates the signal at each relay by detecting the received optical waveform, recovering the transmitted symbol or bit sequence, and retransmitting a newly generated waveform to the next hop. In optical inter-satellite links, this is commonly implemented through optical-to-electrical-to-optical (O/E/O) processing, which converts the received optical signal to the electrical domain for detection, decision making, and decoding or re-encoding before optical retransmission. By making fresh decisions instead of forwarding amplified noisy waveforms, DF suppresses hop-by-hop analog noise accumulation and improves multi-hop reliability. However, O/E/O processing introduces per-hop latency from photodetection, electrical processing, buffering, synchronization, and retransmission, while also increasing relay complexity, power consumption, payload burden, and thermal-management requirements in dense LEO constellations \cite{li2026incrementalhybrid,yahia2022haps}.

In contrast, AF relaying provides transparent forwarding that can be implemented entirely in the optical domain \cite{dabiri2021uavaf}. Rather than detecting or decoding the received signal, an AF relay directly amplifies the incoming optical waveform, typically using erbium-doped fiber amplifiers (EDFAs), and forwards it to the next satellite \cite{cai2019fewmodeedfa}. By avoiding optical-to-electrical (E/O) conversion, digital processing, buffering, and E/O retransmission, all-optical AF offers a low-complexity, low-latency architecture for delay-sensitive LEO satellite links. However, without symbol-level correction, amplified spontaneous emission (ASE), background noise, pointing-induced power fluctuations, and waveform distortion are repeatedly amplified and propagated, causing progressive reliability degradation over long multi-hop optical inter-satellite routes \cite{vu2018allopticaltwoway,nor2017experimental}.

A natural way to bridge the limitations of DF and AF is optical hard-limiter (OHL)-based regeneration, where the relay makes threshold decisions directly on the received optical intensity. Since intensity-modulation/direct-detection (IM/DD) systems carry information in optical power, an OHL can reshape the signal without photodetection, electrical decoding, or E/O retransmission, preserving all-optical low latency while adding regeneration absent in transparent AF relays. However, OHL-assisted relaying remains insufficiently studied for multi-hop optical design. In \cite{trinh2015ohl}, an EDFA-assisted OHL relay for terrestrial FSO links limited background-noise accumulation, but considered binary on-off keying (OOK), atmospheric turbulence, and simplified ASE treatment. In \cite{vu2016twowaync}, OHLs with an all-optical XOR gate were limited to dual-hop binary network-coded FSO relaying and did not model multi-hop error evolution. More recently, \cite{dabiri2025interstellarohl} compared OHL-based inter-satellite relaying with AF and DF, but still assumed binary OOK, simplified EDFA-induced noise, and omitted symbol-level transitions over a regenerated multi-level intensity alphabet. Thus, key questions remain on higher-order OHL regeneration, fixed-threshold behavior under pointing fluctuations, EDFA-induced noise effects, and end-to-end accumulation of relay decision errors.

To address these limitations, we propose an M-PAM all-optical regenerative relay and physical-layer design framework for multi-hop inter-satellite optical networks. Each relay compensates pointing-induced intensity fluctuations through variable-gain amplification, performs threshold-based M-PAM decisions, and regenerates a valid optical power level, thereby retaining all-optical low latency while avoiding AF noise forwarding and DF O/E/O processing. Closed-form per-hop SER and end-to-end transition-matrix models are derived to characterize reliability versus modulation order, threshold scaling, EDFA gain limits, transmit power, pointing jitter, beamwidth, hop count, and optical noise. It therefore offers a design reference for low-delay multi-hop inter-satellite optical relaying. The main contributions of this work are summarized as follows:

\begin{itemize}
\item \textbf{OHL-Assisted All-Optical Regenerative Relaying:} We propose a multi-hop relay architecture using parallel OHL banks for symbol-level regeneration. This eliminates the noise accumulation of AF and the latency-inducing O/E/O conversion and buffering of DF.

\item \textbf{Pointing-Aware EDFA Stabilization:} We develop a variable-gain EDFA mechanism to compensate for pointing-induced fluctuations, maintaining M-PAM level alignment with fixed OHL thresholds while characterizing the resulting gain-limited outage.

\item \textbf{Comprehensive Optical Decision-Statistic Model:} We formulate a relay input statistic by jointly modeling beam-spreading, aperture coupling, pointing misalignment, and all dominant noise sources (ASE-ASE/signal-ASE beat, background, and implementation noise).

\item \textbf{Closed-Form Per-Hop M-PAM SER Analysis:} We derive a tractable closed-form SER expression for general M-PAM that accounts for gain-limited outage, symbol weighting, and signal-ASE/constant-noise regimes, establishing a design benchmark for optical relaying.

\item \textbf{Markov End-to-End SER Framework:} We develop a transition-matrix framework to capture regenerated error propagation across multi-hop routes, enabling closed-form end-to-end SER evaluation for arbitrary modulation and hop counts.

\item \textbf{System-Level Validation and Tradeoff Analysis:} We validate the analytical framework via Monte Carlo (MC) simulations, showing near-exact agreement. Comparisons with AF and DF quantify key parameter impacts, providing actionable design guidelines for low-delay, spectrally efficient, and pointing-aware inter-satellite optical relaying.

\end{itemize}

The remainder of this paper is organized as follows. Section II presents the system model and all-optical relay architecture. Section III develops the pointing-impaired channel and optical noise model. Section IV introduces the variable-gain compensation and OHL-based M-PAM detection rule. Section V derives the per-hop SER. Section VI develops the end-to-end transition-matrix analysis. Section VII presents numerical results. Section VIII concludes the paper.

\section{System Model and Relay Architecture}

\subsection{Multi-Hop Inter-Satellite Topology}

\begin{figure*}[!t]
    \centering
    \includegraphics[width=\textwidth]{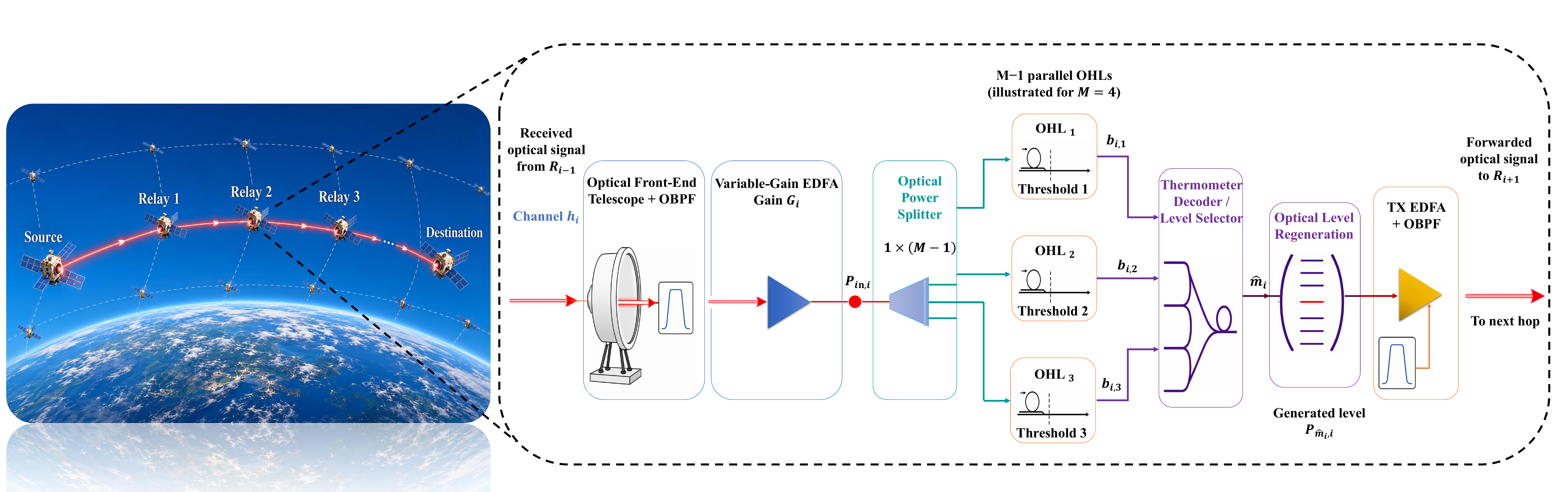}
    \caption{Proposed multi-hop all-optical regenerative M-PAM inter-satellite relay architecture using variable-gain EDFA stabilization and a parallel OHL bank for optical-domain symbol decision and level regeneration.}
    \label{fig:fig1}
\end{figure*}

We consider an all-optical multi-hop inter-satellite optical communication network
in which a source satellite communicates with a destination satellite through a
sequence of intermediate regenerative relay satellites. The communication path is
assumed to be established by a higher-layer routing mechanism; hence, relay
selection and route optimization are outside the scope of this work. The selected
route is represented by the ordered set of nodes
\begin{equation}
\mathcal{R} \triangleq \left\{ R_0, R_1, \ldots, R_{N_r}, R_{N_r+1} \right\},
\label{eq:route_set}
\end{equation}
where $R_0$ denotes the source satellite, $R_{N_r+1}$ denotes the destination
satellite, and $R_i$, $i=1,\ldots,N_r$, denotes the $i$-th intermediate
all-optical regenerative relay. As illustrated in Fig.~\ref{fig:fig1}, information is
forwarded sequentially along the pre-selected relay chain. Let
\begin{equation}
H \triangleq N_r + 1
\label{eq:number_of_hops}
\end{equation}
denote the total number of optical hops between the source and the destination.
The $i$-th hop corresponds to the inter-satellite optical link from $R_{i-1}$
to $R_i$, where $i=1,\ldots,H$. The corresponding hop distance is denoted by
$L_i$, and the total end-to-end propagation distance along the selected route is
\begin{equation}
L_{\mathrm{tot}} = \sum_{i=1}^{H} L_i .
\label{eq:total_distance}
\end{equation}

Here, $L_{\mathrm{tot}}$ is the source-to-destination path length over the selected relay chain, and $L_i$ is the physical distance of the $i$-th inter-satellite hop. Each hop is modeled as a space-based FSO link dominated by geometric spreading and pointing-induced misalignment.

\subsection{Optical M-PAM Signaling Model}
We consider intensity-based M-PAM, where information is encoded in discrete optical power levels. This choice is compatible with the proposed IM/DD all-optical relay because symbol discrimination is performed directly in the optical power domain. The M-PAM symbol alphabet is defined as
\begin{equation}
\mathcal{S} \triangleq \left\{0,1,\ldots,M-1\right\},
\label{eq:symbol_alphabet}
\end{equation}
where $M$ is the modulation order and each symbol $m \in \mathcal{S}$ is
mapped to a unique optical power level. We adopt a uniformly spaced intensity
constellation given by
\begin{equation}
P_m = P_{\min} + m \Delta P,
\qquad
m = 0,1,\ldots,M-1,
\label{eq:mpam_power_levels}
\end{equation}

where $P_m$ is the optical power associated with symbol $m$, $P_{\min} \geq 0$
is the minimum launched optical power, and $\Delta P > 0$ is the constant
spacing between adjacent optical power levels. Let \(m_i \in \mathcal{S}\) denote the symbol regenerated and launched by node \(R_i\), with transmit level \(P_{m_i}\) selected from \eqref{eq:mpam_power_levels}. Each transmitting node therefore launches one of the same \(M\) admissible optical power levels. Unless otherwise stated, the source symbols are assumed equiprobable. Higher-order M-PAM enables increased spectral efficiency while preserving all-optical threshold-based regeneration.

\subsection{Proposed Relay Architecture and Modeling Assumptions}

The proposed relay architecture is shown in Fig.~\ref{fig:fig1}. 
At each intermediate node, the received optical signal is collected by the telescope, filtered by the receive OBPF, and amplified by a receiver-side variable-gain EDFA to stabilize the optical power scale before thresholding. 
The amplified signal is then split into $M-1$ parallel OHL branches, whose outputs form a thermometer pattern used to select one of the admissible $M$-PAM optical levels. 
The selected level is regenerated directly in the optical domain, amplified by a transmit-side EDFA, filtered by a transmit OBPF, and launched toward the next satellite. 
Thus, the relay performs symbol-level optical regeneration without O/E/O conversion, electrical-domain decoding, or buffering. The analysis focuses on physical-layer per-hop and end-to-end SER over a fixed relay path; relay selection and route optimization are outside the scope of this work. Each hop is modeled as a vacuum FSO link with deterministic geometric spreading and random pointing-induced misalignment, while atmospheric and weather-induced impairments are excluded. The system uses intensity modulation, so coherent phase/frequency effects are not considered. Channel gains are independent across hops and follow a block-wise model, remaining approximately constant over one frame; accordingly, the receiver-side EDFA gain is updated at the frame level. After regeneration, each relay decision depends only on the current-hop optical decision statistic, enabling the Markov error-propagation model developed in Section VI.

From a security perspective, the proposed relay performs optical-domain symbol regeneration without electrical decoding or payload processing. Consequently, end-to-end encryption can in principle remain terminated at the source and destination rather than at intermediate relay nodes, which is particularly relevant to multi-vendor operation.

\section{Pointing-Impaired Inter-Satellite Channel and Optical Noise Model}

\subsection{Gaussian Beam Propagation and Aperture Coupling}

We characterize the $i$-th inter-satellite hop, from node $R_{i-1}$ to node $R_i$, where received optical power is affected by Gaussian beam spreading, finite aperture collection, and pointing-induced misalignment. The transmitted beam is modeled as a fundamental Gaussian beam with waist $w_0$. For a propagation distance $L_i$, the beam radius at the receiver plane is given by \cite{farid2007outage}:
\begin{equation}
w_i = w_0 \sqrt{1+\left(\frac{L_i}{z_R}\right)^2},
\label{eq:beam_radius_receiver}
\end{equation}

where $w_i$ is the beam radius at node $R_i$, and $z_R$ is the Rayleigh range
defined as \cite{farid2007outage}:
\begin{equation}
z_R = \frac{\pi w_0^2}{\lambda}.
\label{eq:rayleigh_range}
\end{equation}
In \eqref{eq:rayleigh_range}, $\lambda$ is the optical carrier wavelength. Let $a_i$ denote the radius of the circular receive aperture at node $R_i$. In
the absence of pointing misalignment, the maximum fraction of optical power
collected by the aperture is expressed as \cite{farid2007outage}:
\begin{equation}
A_i =
\left[
\operatorname{erf}
\left(
\frac{\sqrt{\pi}a_i}{\sqrt{2}w_i}
\right)
\right]^2,
\label{eq:aperture_coupling}
\end{equation}

where $\operatorname{erf}(\cdot)$ is the error function and $A_i$,$0 < A_i \leq 1$,, is the deterministic aperture-coupling gain.

\subsection{Pointing-Error Channel-Gain Statistics}

Let $r_i \geq 0$ denote the radial displacement between the beam center and receive-aperture center at node $R_i$. The pointing-impaired channel gain is

\begin{equation}
h_i = A_i \exp\left(-\frac{2r_i^2}{w_i^2}\right),
\qquad
0 < h_i \leq A_i,
\label{eq:pointing_channel_gain}
\end{equation}
where $h_i$ is the instantaneous optical channel gain. The residual angular pointing error of hop $i$ is represented by
$\sigma_{\theta,i}$. The corresponding transverse displacement standard
deviation at the receiver plane is
\begin{equation}
\sigma_{s,i} = L_i \sigma_{\theta,i},
\label{eq:transverse_displacement_std}
\end{equation}

where $\sigma_{s,i}$ is the per-axis spatial jitter standard deviation and
$L_i$ is the hop distance. Accordingly, $r_i$ follows a Rayleigh distribution
with cumulative distribution function \cite{farid2007outage}:
\begin{equation}
F_{r_i}(r)
=
1-\exp\left(-\frac{r^2}{2\sigma_{s,i}^2}\right),
\qquad
r \geq 0.
\label{eq:rayleigh_displacement_cdf}
\end{equation}
Using the monotonic relation between $h_i$ and $r_i$ in
\eqref{eq:pointing_channel_gain}, the cumulative distribution function of
$h_i$ is obtained as
\begin{equation}
F_{h_i}(h)
=
\left(\frac{h}{A_i}\right)^{\xi_i},
\qquad
0 < h \leq A_i,
\label{eq:channel_gain_cdf}
\end{equation}
where the pointing-error severity parameter is
\begin{equation}
\xi_i
=
\frac{w_i^2}{4\sigma_{s,i}^2}
=
\frac{w_i^2}{4L_i^2\sigma_{\theta,i}^2}.
\label{eq:pointing_severity_parameter}
\end{equation}

A larger $\xi_i$ corresponds to weaker pointing fluctuations relative to the
beam footprint. Differentiating \eqref{eq:channel_gain_cdf}, the probability
density function of the channel gain is
\begin{equation}
f_{h_i}(h)
=
\frac{\xi_i}{A_i^{\xi_i}} h^{\xi_i-1},
\qquad
0 < h \leq A_i .
\label{eq:channel_gain_pdf}
\end{equation}

\subsection{Optical Decision Statistic at the Relay Input}

Consider hop $i$, where node $R_{i-1}$ transmits regenerated symbol $m \in \mathcal{S}$ with optical level $P_m$. After transmit-side EDFA amplification, the launched optical power is

\begin{equation}
\widetilde{P}_{m,i-1}
=
G_{i-1}^{\mathrm{TX}} P_m
+
n_{i-1}^{\mathrm{TX}},
\label{eq:tx_edfa_output_power}
\end{equation}
where $\widetilde{P}_{m,i-1}$ is the optical power after transmit-side
amplification, $G_{i-1}^{\mathrm{TX}}$ is the transmit-side EDFA gain at node
$R_{i-1}$, and $n_{i-1}^{\mathrm{TX}}$ denotes the ASE noise generated by that
amplifier.

After propagation with gain $h_i$, the received signal is filtered and amplified by the receiver-side EDFA with gain $G_i$. The OHL-bank input statistic is

\begin{equation}
P_{\mathrm{in},i}
=
G_i h_i \widetilde{P}_{m,i-1}
+
n_i^{\mathrm{RX}}
+
n_i^{\mathrm{bg}}
+
n_i^{\mathrm{th}},
\label{eq:relay_input_decision_statistic}
\end{equation}
where $P_{\mathrm{in},i}$ is the splitter-input optical power,
$n_i^{\mathrm{RX}}$ is receiver-side ASE noise,
$n_i^{\mathrm{bg}}$ is background optical noise after the receive OBPF,
and $n_i^{\mathrm{th}}$ is implementation-dependent decision noise. Substituting \eqref{eq:tx_edfa_output_power} into
\eqref{eq:relay_input_decision_statistic} gives the signal-plus-noise
representation
\begin{equation}
P_{\mathrm{in},i}
=
\mu_{i,m}(h_i) + w_{i,m},
\label{eq:signal_plus_noise_representation}
\end{equation}
where the conditional mean is
\begin{equation}
\mu_{i,m}(h_i)
=
G_i h_i G_{i-1}^{\mathrm{TX}} P_m,
\label{eq:conditional_mean_relay_input}
\end{equation}
and the aggregate noise term is
\begin{equation}
w_{i,m}
=
G_i h_i n_{i-1}^{\mathrm{TX}}
+
n_i^{\mathrm{RX}}
+
n_i^{\mathrm{bg}}
+
n_i^{\mathrm{th}}.
\label{eq:aggregate_noise_relay_input}
\end{equation}
Equations~\eqref{eq:signal_plus_noise_representation}--\eqref{eq:aggregate_noise_relay_input}
define the optical-domain statistic that is later compared with the fixed OHL
thresholds.

\subsection{EDFA ASE, Background, and Implementation Noise}
The noise terms in \eqref{eq:aggregate_noise_relay_input} include transmit-side ASE, receiver-side ASE, background radiation after receive filtering, and implementation-dependent decision noise. For an EDFA stage $x \in \{\mathrm{TX},\mathrm{RX}\}$ with gain $G$, the one-sided ASE PSD is \cite{agrawal2010fiber}:

\begin{equation}
S_{\mathrm{ASE}}^{x}(G)
=
2 n_{\mathrm{sp}}^{x} (G-1) h_P \nu,
\label{eq:ase_psd}
\end{equation}

where $n_{\mathrm{sp}}^{x}$ is the spontaneous-emission factor,
$h_P$ is Planck's constant,
and $\nu$ is the optical carrier frequency. At the decision input, the receiver-side ASE produces an ASE--ASE beat-noise
variance expressed as \cite{kahn2004spectral}:
\begin{equation}
\sigma_{\mathrm{AA},i}^{2}(G_i)
=
\kappa_{\mathrm{AA}}
\left[
S_{\mathrm{ASE}}^{\mathrm{RX}}(G_i)
\right]^2
B_o B_e,
\label{eq:ase_ase_beat_variance}
\end{equation}
where $B_e$ is the equivalent decision bandwidth,
$B_o$ is the optical filter bandwidth,
and $\kappa_{\mathrm{AA}} > 0$ captures implementation, filtering, polarization, and decision-stage effects. The signal--ASE beat-noise variance associated with symbol $m$ is \cite{kahn2004spectral}:
\begin{equation}
\sigma_{\mathrm{SA},i}^{2}(m,h_i)
=
\kappa_{\mathrm{SA}}
\mu_{i,m}(h_i)
S_{\mathrm{ASE}}^{\mathrm{RX}}(G_i)
B_e,
\label{eq:signal_ase_beat_variance}
\end{equation}
where $\sigma_{\mathrm{SA},i}^{2}(m,h_i)$ is the receiver signal--ASE
beat-noise variance, $\mu_{i,m}(h_i)$ is the conditional signal mean in
\eqref{eq:conditional_mean_relay_input}, and $\kappa_{\mathrm{SA}} > 0$ is the
corresponding implementation-dependent coefficient. The transmit-side ASE is also scaled by the channel and receiver-side EDFA, giving \cite{kahn2004spectral}:

\begin{equation}
\sigma_{\mathrm{TX},i}^{2}(h_i)
=
\kappa_{\mathrm{TX}}
\left(G_i h_i\right)^2
\left[
S_{\mathrm{ASE}}^{\mathrm{TX}}
\left(G_{i-1}^{\mathrm{TX}}\right)
\right]^2
B_o B_e,
\label{eq:tx_ase_variance}
\end{equation}
where $\kappa_{\mathrm{TX}} > 0$ captures the corresponding
implementation and filtering factors. The background optical noise and implementation-dependent additive noise are
represented by
\begin{equation}
\sigma_{\mathrm{bg}}^{2}
=
N_{\mathrm{bg}} B_e,
~~~\sigma_{\mathrm{th}}^{2}
=
N_{\mathrm{th}} B_e,
\label{eq:background_noise_variance}
\end{equation}

where $N_{\mathrm{bg}}$ and $N_{\mathrm{th}}$ are the equivalent one-sided PSDs of the background and implementation-dependent decision noises, respectively.

\subsection{Conditional Gaussian Decision Model}

Assuming mutually independent noise components, the aggregate decision-input variance is
\begin{equation}
\sigma_{i,m}^{2}(h_i)
=
\sigma_{\mathrm{bg}}^{2}
+
\sigma_{\mathrm{th}}^{2}
+
\sigma_{\mathrm{AA},i}^{2}(G_i)
+
\sigma_{\mathrm{TX},i}^{2}(h_i)
+
\sigma_{\mathrm{SA},i}^{2}(m,h_i),
\label{eq:total_conditional_noise_variance}
\end{equation}

where the terms are defined in \eqref{eq:ase_ase_beat_variance}--\eqref{eq:background_noise_variance}. Thus, the optical decision statistic follows the conditional Gaussian model \cite{proakis2008digital}:
\begin{equation}
P_{\mathrm{in},i} \mid (m,h_i)
\sim
\mathcal{N}
\left(
\mu_{i,m}(h_i),
\sigma_{i,m}^{2}(h_i)
\right),
\label{eq:conditional_gaussian_decision_model}
\end{equation}

Since both $\mu_{i,m}(h_i)$ and $\sigma_{i,m}^{2}(h_i)$ depend on $G_i$,
the next section develops the gain-control law and all-optical M-PAM detection rule.

\section{Variable-Gain Compensation and All-Optical M-PAM Detection}

\subsection{Threshold Instability With Fixed Optical Hard Limiters}

The OHL bank in relay $R_i$ discriminates symbols by comparing $P_{\mathrm{in},i}$
with fixed equivalent input thresholds $\{\vartheta_{i,k}\}_{k=1}^{M-1}$.
In the absence of noise, each threshold must lie between the conditional means
of two adjacent M-PAM levels. From \eqref{eq:conditional_mean_relay_input}, the conditional mean associated
with symbol $m$ can be written as
\begin{equation}
\mu_{i,m}(h_i)
=
\gamma_i(h_i) P_m,
\label{eq:conditional_mean_scaling}
\end{equation}
where the effective optical scaling factor is
\begin{equation}
\gamma_i(h_i)
\triangleq
G_i h_i G_{i-1}^{\mathrm{TX}} .
\label{eq:effective_optical_scaling_factor}
\end{equation}

 Since the OHL thresholds are fixed in the optical power domain, reliable
slicing between symbols $k-1$ and $k$ requires
\begin{equation}
\gamma_i(h_i) P_{k-1}
<
\vartheta_{i,k}
<
\gamma_i(h_i) P_k,
\qquad
k=1,\ldots,M-1.
\label{eq:reliable_slicing_condition}
\end{equation}

If $G_i$ is fixed while $h_i$ fluctuates, the conditional means in \eqref{eq:conditional_mean_scaling} shift relative to the fixed thresholds, and \eqref{eq:reliable_slicing_condition} may fail even without additive noise. Equivalently, after normalization by $\gamma_i(h_i)$, the
effective decision boundary becomes
\begin{equation}
\bar{\vartheta}_{i,k}(h_i)
=
\frac{\vartheta_{i,k}}{\gamma_i(h_i)},
\qquad
k=1,\ldots,M-1,
\label{eq:normalized_decision_boundary}
\end{equation}
which is random whenever $\gamma_i(h_i)$ is not stabilized. Thus, pointing variations induce random threshold displacement, motivating the variable-gain compensation below.

\subsection{Block-Wise Variable-Gain EDFA Law}

To stabilize \eqref{eq:reliable_slicing_condition}, the receiver-side EDFA gain maintains $\gamma_i(h_i)$
close to a target value $\gamma_0>0$, subject to
$G_{\min}\leq G_i\leq G_{\max}$. Assuming perfect block-wise channel
estimation, the applied gain is
\begin{equation}
G_i(h_i)
=
\left[
\frac{\gamma_0}{h_i G_{i-1}^{\mathrm{TX}}}
\right]_{G_{\min}}^{G_{\max}},
\label{eq:perfect_csi_rx_edfa_gain}
\end{equation}

where $G_{\min}$ and $G_{\max}$ are the minimum and maximum receiver-side
EDFA gains, and the clipping operator is

\begin{equation}
[x]_{G_{\min}}^{G_{\max}}
\triangleq
\min\left\{G_{\max},\max\left\{G_{\min},x\right\}\right\}.
\label{eq:gain_clipping_operator}
\end{equation}

The gain is adapted per frame, while the OHL thresholds remain fixed.

\subsection{Threshold-Stable Region and Gain-Limited Outage}

Substituting \eqref{eq:perfect_csi_rx_edfa_gain} into \eqref{eq:effective_optical_scaling_factor}, the optical scaling becomes

\begin{equation}
\gamma_i(h_i)
=
\begin{cases}
G_{\max} h_i G_{i-1}^{\mathrm{TX}},
&
0 < h_i < h_{L,i},
\\[2mm]
\gamma_0,
&
h_{L,i} \leq h_i \leq h_{U,i},
\\[2mm]
G_{\min} h_i G_{i-1}^{\mathrm{TX}},
&
h_i > h_{U,i},
\end{cases}
\label{eq:piecewise_optical_scaling}
\end{equation}
where the lower and upper gain-transition channel levels are

\begin{equation}
h_{L,i}
=
\frac{\gamma_0}{G_{\max}G_{i-1}^{\mathrm{TX}}},
\qquad
h_{U,i}
=
\frac{\gamma_0}{G_{\min}G_{i-1}^{\mathrm{TX}}}.
\label{eq:gain_transition_levels}
\end{equation}
Here, $h_{L,i}$ is the minimum channel gain for which the receiver EDFA can
realize the target scaling $\gamma_0$, while $h_{U,i}$ is the channel gain above
which the required gain would fall below $G_{\min}$. Therefore, the
threshold-stable operating region is
\begin{equation}
\mathcal{H}_{\mathrm{st},i}
=
\left\{
h_i : h_{L,i} \leq h_i \leq h_{U,i}
\right\}.
\label{eq:threshold_stable_region}
\end{equation}

For $h_i\in\mathcal{H}_{\mathrm{st},i}$, $\gamma_i(h_i)=\gamma_0$ and the
conditional means align with the fixed thresholds. For $h_i<h_{L,i}$, the
EDFA saturates at $G_{\max}$, defining the gain-limited outage event
$\mathcal{O}_i \triangleq \{h_i<h_{L,i}\}$. Using the channel-gain CDF in \eqref{eq:channel_gain_cdf}, the corresponding
per-hop outage probability is

\begin{equation}
P_{\mathrm{out},i}
=
\Pr\{\mathcal{O}_i\}
=
\left(
\frac{h_{L,i}}{A_i}
\right)^{\xi_i},
\qquad
0 < h_{L,i} \leq A_i .
\label{eq:per_hop_outage_probability}
\end{equation}

\subsection{Parallel OHL Bank and Thermometer Decoding}

At relay $R_i$, $P_{\mathrm{in},i}$ is split into $M-1$ parallel OHL branches. Let $\eta_{i,k}$ denote the optical splitting coefficient
of branch $k$, where
\begin{equation}
0 < \eta_{i,k} < 1,
\qquad
k=1,\ldots,M-1,
\qquad
\sum_{k=1}^{M-1} \eta_{i,k} \leq 1.
\label{eq:splitter_coefficient_range}
\end{equation}
The inequality in \eqref{eq:splitter_coefficient_range} accounts for splitter
insertion loss and, if required, an auxiliary monitoring tap. The optical power
entering the $k$-th OHL branch is
\begin{equation}
P_{i,k}^{\mathrm{br}}
=
\eta_{i,k} P_{\mathrm{in},i},
\qquad
k=1,\ldots,M-1,
\label{eq:ohl_branch_input_power}
\end{equation}

 Each OHL
compares this branch power with a fixed physical threshold $\tau_{i,k}>0$.
Since $\eta_{i,k}>0$, the binary output of the $k$-th OHL can be equivalently written as

\begin{equation}
\begin{aligned}
b_{i,k}
&= \mathbf{1}\!\left\{P_{i,k}^{\mathrm{br}} \geq \tau_{i,k}\right\}
= \mathbf{1}\!\left\{P_{\mathrm{in},i} \geq \theta_{i,k}\right\},\\
&\qquad k=1,\ldots,M-1 .
\end{aligned}
\label{eq:ohl_binary_output_branch}
\end{equation}

where $b_{i,k}\in\{0,1\}$, and $\mathbf{1}\{\cdot\}$ denotes the indicator
function, $\theta_{i,k}
\triangleq
\frac{\tau_{i,k}}{\eta_{i,k}}$ is the equivalent input threshold of the $k$-th OHL. The
thresholds are selected to satisfy the ordering

\begin{equation}
0 < \theta_{i,1} < \theta_{i,2} < \cdots < \theta_{i,M-1}.
\label{eq:ordered_equivalent_thresholds}
\end{equation}
The relay decision is obtained by counting the exceeded thresholds:

\begin{equation}
\widehat{m}_i
=
\sum_{k=1}^{M-1} b_{i,k},
\label{eq:thermometer_decoding_rule}
\end{equation}

With $\theta_{i,0}\triangleq -\infty$ and $\theta_{i,M}\triangleq +\infty$,
the detector is equivalently

\begin{equation}
\widehat{m}_i = b
\quad \Longleftrightarrow \quad
\theta_{i,b}
\leq
P_{\mathrm{in},i}
<
\theta_{i,b+1},
\qquad
b=0,\ldots,M-1.
\label{eq:interval_decision_rule}
\end{equation}

Threshold placement under the target scaling $\gamma_0$ is specified next.

\subsection{Midpoint Threshold Design and Optical Regeneration}

In the threshold-stable region, $\gamma_i(h_i)=\gamma_0$, so

\begin{equation}
\mu_{i,m}(h_i)
=
\gamma_0 P_m,
\qquad
h_i \in \mathcal{H}_{\mathrm{st},i}.
\label{eq:stable_region_conditional_mean}
\end{equation}
The $k$-th equivalent input threshold is placed between the stabilized means of
symbols $k-1$ and $k$. Using midpoint slicing, we set
\begin{equation}
\theta_{i,k}
=
\frac{\gamma_0 P_{k-1}+\gamma_0 P_k}{2},
\qquad
k=1,\ldots,M-1.
\label{eq:midpoint_threshold_general}
\end{equation}
Substituting the M-PAM power levels from
\eqref{eq:mpam_power_levels}, \eqref{eq:midpoint_threshold_general} becomes
\begin{equation}
\theta_{i,k}
=
\gamma_0
\left[
P_{\min}
+
\left(k-\frac{1}{2}\right)\Delta P
\right],
\qquad
k=1,\ldots,M-1.
\label{eq:midpoint_threshold_mpam}
\end{equation}

The corresponding physical OHL threshold in branch $k$ is 
$\tau_{i,k}
=
\eta_{i,k}\theta_{i,k}$. After thermometer decoding, relay $R_i$ regenerates the detected symbol
$\widehat{m}_i$ into the optical power level
\begin{equation}
P_{\widehat{m}_i}
=
P_{\min}
+
\widehat{m}_i\Delta P.
\label{eq:regenerated_detected_power}
\end{equation}
Thus, the relay output is constrained to the same M-PAM intensity alphabet
defined in \eqref{eq:mpam_power_levels}. The regenerated symbol then becomes
the transmitted symbol for the next hop, while the analog noise realization
that affected the current decision is not forwarded as a continuous analog
waveform. This completes the all-optical detection and regeneration model used
in the per-hop error analysis.

\section{Per-Hop Symbol Error Probability Analysis}

\subsection{Exact Conditional Decision Probabilities}

For hop $i$, suppose that node $R_{i-1}$ regenerates and transmits symbol
$m \in \mathcal{S}$, and the instantaneous channel gain is $h_i=h$. The OHL
detector at node $R_i$ assigns the received statistic $P_{\mathrm{in},i}$ to
symbol $b \in \mathcal{S}$ according to the interval rule in
\eqref{eq:interval_decision_rule}. Using the conditional Gaussian model in
\eqref{eq:conditional_gaussian_decision_model}, the probability of deciding
symbol $b$ when symbol $m$ was transmitted is

\begin{equation}
\begin{aligned}
p_{i,b|m}(h)
&\triangleq
\Pr\left\{
\widehat{m}_i=b \mid m, h_i=h
\right\} \\
&=
\Phi
\left(
\frac{\theta_{i,b+1}-\mu_{i,m}(h)}
{\sigma_{i,m}(h)}
\right)
-
\Phi
\left(
\frac{\theta_{i,b}-\mu_{i,m}(h)}
{\sigma_{i,m}(h)}
\right).
\end{aligned}
\label{eq:conditional_decision_probability}
\end{equation}

where $p_{i,b|m}(h)$ is the conditional symbol decision probability, $\Phi(\cdot)$ denote the standard
Gaussian CDF, $\widehat{m}_i$ is the symbol detected at relay $R_i$, $\theta_{i,b}$ and
$\theta_{i,b+1}$ are the lower and upper decision thresholds associated with
output symbol $b$, $\mu_{i,m}(h)$ is the conditional mean in
\eqref{eq:conditional_mean_relay_input}, and $\sigma_{i,m}(h)$ is the square
root of the conditional variance in
\eqref{eq:total_conditional_noise_variance}. The corresponding conditional probability of symbol error for transmitted
symbol $m$ is
\begin{equation}
P_{e,i}(m|h)
=
1-p_{i,m|m}(h),
\label{eq:conditional_symbol_error_probability}
\end{equation}
where $P_{e,i}(m|h)$ denotes the probability that relay $R_i$ regenerates a
symbol different from $m$, conditioned on $h_i=h$. Equation~\eqref{eq:conditional_decision_probability}
is valid for all output symbols, including the two edge symbols.

\subsection{Conditional and Average Per-Hop SER}

Using \eqref{eq:conditional_symbol_error_probability}, the non-outage
conditional SER of hop $i$ is obtained by averaging over the equiprobable
M-PAM symbols as
\begin{equation}
P_{s,i}^{\mathrm{nl}}(h)
=
\frac{1}{M}
\sum_{m=0}^{M-1}
P_{e,i}(m|h),
\qquad
h_{L,i} \leq h \leq A_i,
\label{eq:non_outage_conditional_ser}
\end{equation}
where $P_{s,i}^{\mathrm{nl}}(h)$ denotes the conditional SER outside the
gain-limited outage region. In
\eqref{eq:non_outage_conditional_ser}, the quantities
$P_{e,i}(m|h)$, $\mu_{i,m}(h)$, and $\sigma_{i,m}(h)$ are evaluated using
\eqref{eq:conditional_symbol_error_probability},
\eqref{eq:conditional_mean_relay_input}, and
\eqref{eq:total_conditional_noise_variance}, respectively, with the gain
$G_i(h)$ given by \eqref{eq:perfect_csi_rx_edfa_gain}. For $0<h<h_{L,i}$, the receiver-side EDFA saturates at $G_{\max}$ and the
target scaling $\gamma_0$ cannot be achieved. In this deep-fade regime, the
detector output is modeled as being dominated by the lowest optical level.
Therefore, the outage-region SER is

\begin{equation}
P_{s,i}^{\mathrm{out}}
=
\frac{M-1}{M},
\label{eq:outage_region_ser}
\end{equation}
The overall conditional SER is
\begin{equation}
P_{s,i}(h)
=
\begin{cases}
P_{s,i}^{\mathrm{out}},
&
0<h<h_{L,i},
\\[1mm]
P_{s,i}^{\mathrm{nl}}(h),
&
h_{L,i}\leq h\leq A_i .
\end{cases}
\label{eq:overall_conditional_ser}
\end{equation}
Averaging \eqref{eq:overall_conditional_ser} over the pointing-induced channel
gain yields the exact per-hop average SER
\begin{equation}
\overline{P}_{s,i}
=
P_{s,i}^{\mathrm{out}} P_{\mathrm{out},i}
+
\frac{\xi_i}{A_i^{\xi_i}}
\int_{h_{L,i}}^{A_i}
h^{\xi_i-1}
P_{s,i}^{\mathrm{nl}}(h)
\,dh .
\label{eq:average_ser_expanded_pdf}
\end{equation}

where $P_{\mathrm{out},i}$ is given in
\eqref{eq:per_hop_outage_probability}. If $h_{L,i}>A_i$, the gain target is unattainable over the full channel support
and $\overline{P}_{s,i}=P_{s,i}^{\mathrm{out}}$. Otherwise,
\eqref{eq:average_ser_expanded_pdf} provides the exact numerical per-hop SER
benchmark before introducing any closed-form approximation.

\subsection{Signal--ASE-Dominant Two-Region Closed Form}

The exact average SER in \eqref{eq:average_ser_expanded_pdf} can be evaluated
numerically using the full conditional variance in
\eqref{eq:total_conditional_noise_variance}. To obtain an analytical
expression, we consider the gain-controlled non-outage region in which
$h_{L,i}\leq h \leq A_i$ and the target scaling $\gamma_i(h)=\gamma_0$ is
maintained. This corresponds to the design condition $h_{U,i}\geq A_i$; if this
condition is not satisfied, the exact average SER in
\eqref{eq:average_ser_expanded_pdf} should be used. The Gaussian $Q$-function is defined as
\begin{equation}
Q(x)
=
\frac{1}{\sqrt{2\pi}}
\int_{x}^{\infty}
\exp\left(-\frac{t^2}{2}\right)
\,dt .
\label{eq:q_function_definition}
\end{equation}
Under midpoint thresholding, the distance between each stabilized conditional
mean and its nearest decision threshold is
\begin{equation}
d_0
=
\frac{\gamma_0 \Delta P}{2},
\label{eq:nearest_threshold_distance}
\end{equation}

where $d_0$ is the half-distance between adjacent stabilized M-PAM levels at the
OHL input. The edge and interior symbol weights are
\begin{equation}
w_m
=
\begin{cases}
1,
&
m=0 \ \text{or} \ m=M-1,
\\[1mm]
2,
&
m=1,\ldots,M-2,
\end{cases}
\label{eq:edge_interior_symbol_weights}
\end{equation}
where $w_m$ accounts for the number of nearest decision boundaries associated
with symbol $m$. Therefore, using the outage decomposition in
\eqref{eq:average_ser_expanded_pdf}, the one-hop average SER can be written as
\begin{equation}
\begin{aligned}
\overline{P}_{s,i}
&=
\frac{M-1}{M}
\left(
\frac{h_{L,i}}{A_i}
\right)^{\xi_i}
\\
&\quad
+
\frac{\xi_i}{M A_i^{\xi_i}}
\sum_{m=0}^{M-1}
w_m
\int_{h_{L,i}}^{A_i}
h^{\xi_i-1}
Q
\left(
\frac{d_0}{\sigma_{i,m}(h)}
\right)
dh .
\end{aligned}
\label{eq:one_hop_average_ser_qform}
\end{equation}
In the considered non-outage region, the receiver-side gain is
\begin{equation}
G_i(h)
=
\frac{\gamma_0}{hG_{i-1}^{\mathrm{TX}}},
\qquad
h_{L,i}\leq h\leq A_i .
\label{eq:non_outage_rx_gain}
\end{equation}

Substituting \eqref{eq:non_outage_rx_gain} into the receiver signal--ASE
beat-noise term gives the following tractable variance model:
\begin{equation}
\sigma_{i,m}^{2}(h)
\approx
\sigma_{\mathrm{eff},i,m}^{2}
+
\frac{D_{i,m}}{h},
\qquad
h_{L,i}\leq h \leq A_i,
\label{eq:tractable_variance_model}
\end{equation}
where
\begin{equation}
D_{i,m}
=
\kappa_{\mathrm{SA}}
\left(\gamma_0 P_m\right)
\left(
\frac{2n_{\mathrm{sp}}^{\mathrm{RX}} h_P \nu \gamma_0}
{G_{i-1}^{\mathrm{TX}}}
\right)
B_e,
\label{eq:signal_ase_coefficient_D}
\end{equation}
and
\begin{equation}
\sigma_{\mathrm{eff},i,m}^{2}
=
\sigma_{\mathrm{bg}}^{2}
+
\sigma_{\mathrm{th}}^{2}
-
\kappa_{\mathrm{SA}}
\left(\gamma_0 P_m\right)
\left(2n_{\mathrm{sp}}^{\mathrm{RX}}h_P\nu\right)
B_e .
\label{eq:effective_noise_floor}
\end{equation}
Here, $D_{i,m}/h$ is the channel-dependent signal--ASE variance component,
while $\sigma_{\mathrm{eff},i,m}^{2}$ is the effective channel-independent
noise floor. The negative correction term in
\eqref{eq:effective_noise_floor} appears because the receiver ASE PSD contains
$G_i(h)-1$, not only $G_i(h)$. Under the high-gain approximation
$G_i(h)-1\simeq G_i(h)$, this correction is negligible and
$\sigma_{\mathrm{eff},i,m}^{2}\simeq \sigma_{\mathrm{bg}}^{2}
+\sigma_{\mathrm{th}}^{2}$.

Define the symbol-dependent integral
\begin{equation}
J_{i,m}
\triangleq
\int_{h_{L,i}}^{A_i}
h^{\xi_i-1}
Q
\left(
\frac{d_0}
{\sqrt{\sigma_{\mathrm{eff},i,m}^{2}+D_{i,m}/h}}
\right)
dh .
\label{eq:symbol_dependent_integral}
\end{equation}
The switching boundary between the signal--ASE-dominant region and the
constant-noise-dominant region is obtained from
$D_{i,m}/h=\sigma_{\mathrm{eff},i,m}^{2}$, yielding
\begin{equation}
h_{i,m}^{\star}
=
\frac{D_{i,m}}{\sigma_{\mathrm{eff},i,m}^{2}},
\qquad
\sigma_{\mathrm{eff},i,m}^{2}>0.
\label{eq:switching_boundary}
\end{equation}
The corresponding clipped integration bounds are
\begin{equation}
\begin{aligned}
b_{1,i,m}
&=
\min\left\{A_i,h_{i,m}^{\star}\right\},
\\
b_{2,i,m}
&=
\max\left\{h_{L,i},h_{i,m}^{\star}\right\}.
\end{aligned}
\label{eq:clipped_integration_bounds}
\end{equation}

Thus, the variance approximation is partitioned as
\begin{equation}
\sigma_{i,m}^{2}(h)
\approx
\begin{cases}
\dfrac{D_{i,m}}{h},
&
h_{L,i}\leq h\leq b_{1,i,m},
\\[3mm]
\sigma_{\mathrm{eff},i,m}^{2},
&
b_{2,i,m}\leq h\leq A_i .
\end{cases}
\label{eq:two_region_variance_approximation}
\end{equation}
Using \eqref{eq:two_region_variance_approximation}, the integral in
\eqref{eq:symbol_dependent_integral} is approximated by
\begin{equation}
\begin{aligned}
J_{i,m}
&\approx
\int_{h_{L,i}}^{b_{1,i,m}}
h^{\xi_i-1}
Q
\left(
\sqrt{\frac{d_0^2}{D_{i,m}}h}
\right)
dh
\\
&\quad
+
\int_{b_{2,i,m}}^{A_i}
h^{\xi_i-1}
Q
\left(
\frac{d_0}{\sigma_{\mathrm{eff},i,m}}
\right)
dh .
\end{aligned}
\label{eq:two_region_integral_approximation}
\end{equation}
For the signal--ASE-dominant part, define the change of variable
\begin{equation}
t
=
\frac{d_0^2}{2D_{i,m}}h,
\label{eq:signal_ase_change_variable}
\end{equation}

with the corresponding bounds
\begin{equation}
t_{L,i,m}
=
\frac{d_0^2 h_{L,i}}{2D_{i,m}},
\qquad
t_{1,i,m}
=
\frac{d_0^2 b_{1,i,m}}{2D_{i,m}}.
\label{eq:signal_ase_t_bounds}
\end{equation}
Using $Q(x)=\frac{1}{2}\operatorname{erfc}(x/\sqrt{2})$, the
signal--ASE-dominant integral admits the closed form
\begin{align}
&\int_{h_{L,i}}^{b_{1,i,m}}
h^{\xi_i-1}
Q
\left(
\sqrt{\frac{d_0^2}{D_{i,m}}h}
\right)
dh
\nonumber\\
&\quad =
\frac{1}{\xi_i}
\left(
\frac{2D_{i,m}}{d_0^2}
\right)^{\xi_i}
\frac{1}{2\sqrt{\pi}}
\left[
\Phi_{\xi_i}(t_{1,i,m})
-
\Phi_{\xi_i}(t_{L,i,m})
\right],
\label{eq:signal_ase_integral_closed_form}
\end{align}
where
\begin{equation}
\Phi_{\xi}(t)
\triangleq
t^{\xi}
\Gamma\left(\frac{1}{2},t\right)
-
\Gamma\left(\xi+\frac{1}{2},t\right),
\label{eq:phi_xi_definition}
\end{equation}

and $\Gamma(a,t)$ is the upper incomplete gamma function. Substituting the
bounds explicitly, \eqref{eq:signal_ase_integral_closed_form} becomes
\begin{align}
&\int_{h_{L,i}}^{b_{1,i,m}}
h^{\xi_i-1}
Q
\left(
\sqrt{\frac{d_0^2}{D_{i,m}}h}
\right)
dh
\nonumber\\
&\quad =
\frac{1}{\xi_i}
\left(
\frac{2D_{i,m}}{d_0^2}
\right)^{\xi_i}
\frac{1}{2\sqrt{\pi}}
\Bigg[
\left(
\frac{d_0^2 b_{1,i,m}}{2D_{i,m}}
\right)^{\xi_i}
\Gamma
\left(
\frac{1}{2},
\frac{d_0^2 b_{1,i,m}}{2D_{i,m}}
\right)
\nonumber\\
&\qquad
-
\Gamma
\left(
\xi_i+\frac{1}{2},
\frac{d_0^2 b_{1,i,m}}{2D_{i,m}}
\right)
\nonumber\\
&\qquad
-
\left(
\frac{d_0^2 h_{L,i}}{2D_{i,m}}
\right)^{\xi_i}
\Gamma
\left(
\frac{1}{2},
\frac{d_0^2 h_{L,i}}{2D_{i,m}}
\right)
\nonumber\\
&\qquad
+
\Gamma
\left(
\xi_i+\frac{1}{2},
\frac{d_0^2 h_{L,i}}{2D_{i,m}}
\right)
\Bigg].
\label{eq:signal_ase_integral_explicit_closed_form}
\end{align}
For the constant-noise-dominant region, the integral is elementary:
\begin{equation}
\int_{b_{2,i,m}}^{A_i}
h^{\xi_i-1}
Q
\left(
\frac{d_0}{\sigma_{\mathrm{eff},i,m}}
\right)
dh
=
Q
\left(
\frac{d_0}{\sigma_{\mathrm{eff},i,m}}
\right)
\frac{A_i^{\xi_i}-b_{2,i,m}^{\xi_i}}{\xi_i}.
\label{eq:constant_noise_integral_closed_form}
\end{equation}

Combining \eqref{eq:signal_ase_integral_explicit_closed_form} and
\eqref{eq:constant_noise_integral_closed_form}, define the two-region
closed-form approximation of $J_{i,m}$ as
\begin{align}
J_{i,m}^{\mathrm{pw}}
&\triangleq
\frac{1}{\xi_i}
\left(
\frac{2D_{i,m}}{d_0^2}
\right)^{\xi_i}
\frac{1}{2\sqrt{\pi}}
\Bigg[
\left(
\frac{d_0^2 b_{1,i,m}}{2D_{i,m}}
\right)^{\xi_i}
\Gamma
\left(
\frac{1}{2},
\frac{d_0^2 b_{1,i,m}}{2D_{i,m}}
\right)
\nonumber\\
&\qquad
-
\Gamma
\left(
\xi_i+\frac{1}{2},
\frac{d_0^2 b_{1,i,m}}{2D_{i,m}}
\right)
\nonumber\\
&\qquad
-
\left(
\frac{d_0^2 h_{L,i}}{2D_{i,m}}
\right)^{\xi_i}
\Gamma
\left(
\frac{1}{2},
\frac{d_0^2 h_{L,i}}{2D_{i,m}}
\right)
\nonumber\\
&\qquad
+
\Gamma
\left(
\xi_i+\frac{1}{2},
\frac{d_0^2 h_{L,i}}{2D_{i,m}}
\right)
\Bigg]
\nonumber\\
&\qquad
+
Q
\left(
\frac{d_0}{\sigma_{\mathrm{eff},i,m}}
\right)
\frac{A_i^{\xi_i}-b_{2,i,m}^{\xi_i}}{\xi_i}.
\label{eq:two_region_closed_form_integral}
\end{align}
Substituting $J_{i,m}^{\mathrm{pw}}$ into
\eqref{eq:one_hop_average_ser_qform}, the compact two-region closed-form
average SER is
\begin{equation}
\overline{P}_{s,i}^{\mathrm{pw}}
=
\frac{M-1}{M}
\left(
\frac{h_{L,i}}{A_i}
\right)^{\xi_i}
+
\frac{\xi_i}{M A_i^{\xi_i}}
\sum_{m=0}^{M-1}
w_m J_{i,m}^{\mathrm{pw}}.
\label{eq:two_region_closed_form_average_ser}
\end{equation}

The compact expression in \eqref{eq:two_region_closed_form_average_ser}, together with
$J_{i,m}^{\mathrm{pw}}$ in \eqref{eq:two_region_closed_form_integral}, provide the final closed-form per-hop average SER under the signal–ASE-dominant plus constant-noise two-region approximation. This formulation preserves the gain-limited outage term, the M-PAM edge/interior symbol weighting, and the explicit dependence of the receiver signal–ASE noise on the gain-control law and pointing-induced channel gain, where $d_0$, $D_{i,m}$,
$\sigma_{\mathrm{eff},i,m}^{2}$, and the bounds $b_{1,i,m}$ and
$b_{2,i,m}$ are defined in \eqref{eq:nearest_threshold_distance}, \eqref{eq:signal_ase_coefficient_D},
\eqref{eq:effective_noise_floor}, and \eqref{eq:clipped_integration_bounds}, respectively.

\section{End-to-End Multi-Hop Error Propagation}

\subsection{Markov State Representation}

The proposed relay chain performs regeneration at every intermediate node;
hence, the end-to-end error behavior is governed by the evolution of discrete
symbol decisions across the $H$ hops. Let
\begin{equation}
X_i \in \mathcal{S},
\qquad
i=0,1,\ldots,H,
\label{eq:markov_symbol_state}
\end{equation}
denote the symbol index regenerated at node $R_i$, where $X_0$ is the source
symbol and $X_H$ is the final symbol detected at the destination. Since
$\mathcal{S}$ was defined in \eqref{eq:symbol_alphabet}, the state space of the
end-to-end process is the same $M$-ary intensity alphabet used by all relay
nodes. For hop $i$, the symbol $X_i$ depends on $X_{i-1}$ through the local channel
realization, optical noise, gain control, and OHL decision rule of that hop.
After regeneration, previous analog noise samples are not forwarded as waveform
components. Therefore, conditioned on the regenerated input symbol of the
current hop, the next regenerated symbol is independent of earlier relay
decisions. This gives the Markov property
\begin{equation}
\begin{aligned}
&\Pr\left\{
X_i=b \mid X_{i-1}=a,X_{i-2},\ldots,X_0
\right\}
 \\
&\quad
=\Pr\left\{
X_i=b \mid X_{i-1}=a
\right\}.
\end{aligned}
\label{eq:markov_property}
\end{equation}
for $a,b\in\mathcal{S}$ and $i=1,\ldots,H$. Thus,
$\{X_i\}_{i=0}^{H}$ forms a finite-state Markov chain whose states are the
regenerated M-PAM symbols. Let $\boldsymbol{\pi}_i \in \mathbb{R}^{1\times M}$ denote the row vector of
symbol probabilities after node $R_i$, with entries
\begin{equation}
[\boldsymbol{\pi}_i]_b
=
\Pr\{X_i=b\},
\qquad
b=0,\ldots,M-1.
\label{eq:state_probability_vector_entries}
\end{equation}
The initial state distribution is determined by the source symbols. Under the
equiprobable assumption
\begin{equation}
\boldsymbol{\pi}_0
=
\frac{1}{M}\mathbf{1}_{1\times M},
\label{eq:initial_state_distribution}
\end{equation}
where $\mathbf{1}_{1\times M}$ is the all-one row vector of length $M$. The
per-hop transition matrix that governs the recursion of $\boldsymbol{\pi}_i$ is
derived next.

\subsection{Per-Hop Transition Matrix}

For hop $i$, define the transition matrix $\mathbf{T}_i\in\mathbb{R}^{M\times M}$
by
\begin{equation}
[\mathbf{T}_i]_{a,b}
\triangleq
\Pr\{X_i=b \mid X_{i-1}=a\},
\qquad
a,b\in\mathcal{S}.
\label{eq:per_hop_transition_matrix}
\end{equation}
Here, $[\mathbf{T}_i]_{a,b}$ is the probability that node $R_i$ regenerates
symbol $b$ when node $R_{i-1}$ has transmitted symbol $a$. With this definition,
the state distribution evolves as
\begin{equation}
\boldsymbol{\pi}_i
=
\boldsymbol{\pi}_{i-1}\mathbf{T}_i,
\qquad
i=1,\ldots,H.
\label{eq:state_distribution_recursion}
\end{equation}
The transition matrix includes both the gain-limited outage region and the
non-outage detection region. Using the outage event in
\eqref{eq:per_hop_outage_probability}, we decompose
\begin{equation}
\mathbf{T}_i
=
P_{\mathrm{out},i}\mathbf{T}_i^{\mathrm{out}}
+
\mathbf{T}_i^{\mathrm{nl}},
\label{eq:transition_matrix_decomposition}
\end{equation}
where $\mathbf{T}_i^{\mathrm{out}}$ is the transition matrix conditioned on
gain-limited outage, and $\mathbf{T}_i^{\mathrm{nl}}$ is the direct non-outage
contribution obtained by averaging over $h_i\in[h_{L,i},A_i]$ using the original
channel-gain PDF. In the outage region, the target optical scaling cannot be reached and the OHL
bank is modeled as collapsing to the lowest regenerated optical level.
Therefore,
\begin{equation}
[\mathbf{T}_i^{\mathrm{out}}]_{a,b}
=
\mathbf{1}\{b=0\},
\qquad
a,b\in\mathcal{S}.
\label{eq:outage_transition_matrix}
\end{equation}
For the non-outage region, the exact conditional decision probability is given
by \eqref{eq:conditional_decision_probability}. Hence, the non-outage
contribution is
\begin{equation}
[\mathbf{T}_i^{\mathrm{nl}}]_{a,b}
=
\int_{h_{L,i}}^{A_i}
p_{i,b|a}(h) f_{h_i}(h)\,dh,
\qquad
a,b\in\mathcal{S},
\label{eq:non_outage_transition_matrix}
\end{equation}
where $p_{i,b|a}(h)$ is the probability that relay $R_i$ decides symbol $b$
when symbol $a$ was transmitted over hop $i$, and $f_{h_i}(h)$ is given in
\eqref{eq:channel_gain_pdf}. Equivalently,
\begin{equation}
[\mathbf{T}_i^{\mathrm{nl}}]_{a,b}
=
\frac{\xi_i}{A_i^{\xi_i}}
\int_{h_{L,i}}^{A_i}
h^{\xi_i-1}
p_{i,b|a}(h)\,dh.
\label{eq:non_outage_transition_matrix_expanded}
\end{equation}

Since \eqref{eq:non_outage_transition_matrix} integrates over the original,
non-normalized channel density on $[h_{L,i},A_i]$, the row sum of
$\mathbf{T}_i^{\mathrm{nl}}$ is $1-P_{\mathrm{out},i}$. Consequently,
\begin{equation}
\sum_{b=0}^{M-1}
[\mathbf{T}_i]_{a,b}
=
1,
\qquad
a\in\mathcal{S},
\label{eq:transition_matrix_row_sum}
\end{equation}
and $\mathbf{T}_i$ is a valid stochastic matrix. If $h_{L,i}>A_i$, then
$P_{\mathrm{out},i}=1$ and $\mathbf{T}_i=\mathbf{T}_i^{\mathrm{out}}$.
Otherwise, \eqref{eq:transition_matrix_decomposition}--\eqref{eq:non_outage_transition_matrix_expanded}
provide the exact per-hop transition model before applying the two-region
closed-form approximation.

\subsection{Closed-Form Non-Outage Transition Entries}

We now express the non-outage contribution in
\eqref{eq:non_outage_transition_matrix_expanded} using the two-region variance
model introduced in Section~V-C. Under midpoint thresholding and target scaling,
the difference between the $k$-th threshold and the conditional mean associated
with transmitted symbol $a$ is
\begin{equation}
\begin{aligned}
\Delta_{a,k}
&\triangleq
\theta_{i,k}-\mu_{i,a}(h)
\\
&=
\gamma_0 \Delta P
\left(
k-a-\frac{1}{2}
\right),
\qquad
k=1,\ldots,M-1 .
\end{aligned}
\label{eq:threshold_mean_offset}
\end{equation}
Here, $\Delta_{a,k}$ is independent of $h$ because the non-outage region
maintains $\gamma_i(h)=\gamma_0$. For $D_{i,a}>0$ and
$\sigma_{\mathrm{eff},i,a}^{2}>0$, define
\begin{equation}
\alpha_{i,a,k}
\triangleq
\frac{\Delta_{a,k}}{\sqrt{D_{i,a}}},
\qquad
\beta_{i,a,k}
\triangleq
\frac{\Delta_{a,k}}{\sigma_{\mathrm{eff},i,a}},
\label{eq:alpha_beta_transition_parameters}
\end{equation}
where $D_{i,a}$ and $\sigma_{\mathrm{eff},i,a}^{2}$ are obtained from \eqref{eq:signal_ase_coefficient_D} and \eqref{eq:effective_noise_floor} by
setting $m=a$, and
$\sigma_{\mathrm{eff},i,a}\triangleq\sqrt{\sigma_{\mathrm{eff},i,a}^{2}}$. For compact notation, define
\begin{equation}
\mathcal{P}_{\nu}(u,v)
\triangleq
\begin{cases}
\dfrac{v^{\nu}-u^{\nu}}{\nu},
&
v>u,
\\[2mm]
0,
&
v\leq u,
\end{cases}
\label{eq:power_interval_function}
\end{equation}
where $\nu>0$. Define the auxiliary function
\begin{equation}
\Omega_{\nu}(t)
\triangleq
t^{\nu}\Gamma\!\left(\frac{1}{2},t\right)
-
\Gamma\!\left(\nu+\frac{1}{2},t\right),
\qquad
t\ge 0.
\label{eq:neweq}
\end{equation}
Using this auxiliary function, define also

\begin{equation}
\begin{aligned}
\mathcal{Q}_{\nu}(s;u,v)
&\triangleq
\frac{1}{\nu}
\left(
\frac{2}{s^2}
\right)^{\nu}
\frac{
\Omega_{\nu}\left(\dfrac{s^2v}{2}\right)
-
\Omega_{\nu}\left(\dfrac{s^2u}{2}\right)
}
{2\sqrt{\pi}},
\\
&\qquad
s>0,\; v>u .
\end{aligned}
\label{eq:q_interval_function}
\end{equation}
For $v\leq u$, $\mathcal{Q}_{\nu}(s;u,v)$ is set to zero. The integral of the
Gaussian CDF over the signal--ASE-dominant region is then written as
\begin{align}
\mathcal{F}_{\nu}(\alpha;u,v)
&\triangleq
\int_{u}^{v}
h^{\nu-1}
\Phi\left(\alpha\sqrt{h}\right)
dh
\nonumber\\
&=
\begin{cases}
\mathcal{P}_{\nu}(u,v)-\mathcal{Q}_{\nu}(\alpha;u,v),
&
\alpha>0,
\\[1mm]
\dfrac{1}{2}\mathcal{P}_{\nu}(u,v),
&
\alpha=0,
\\[2mm]
\mathcal{Q}_{\nu}(|\alpha|;u,v),
&
\alpha<0.
\end{cases}
\label{eq:gaussian_cdf_signal_ase_integral}
\end{align}
Similarly, the constant-noise-region integral is
\begin{equation}
\mathcal{C}_{\nu}(c;u,v)
\triangleq
\Phi(c)\mathcal{P}_{\nu}(u,v),
\label{eq:constant_noise_cdf_integral}
\end{equation}
where $c$ is a normalized threshold offset. Using the clipped bounds $b_{1,i,a}$ and $b_{2,i,a}$ from
\eqref{eq:clipped_integration_bounds}, the closed-form non-outage transition
entries are obtained as follows. For $b=0$,
\begin{equation}
\begin{aligned}
[\mathbf{T}_{i}^{\mathrm{nl,pw}}]_{a,0}
&=
\frac{\xi_i}{A_i^{\xi_i}}
\Bigg[
\mathcal{F}_{\xi_i}
\left(
\alpha_{i,a,1};
h_{L,i}, b_{1,i,a}
\right)
\\
&\qquad\quad
+
\mathcal{C}_{\xi_i}
\left(
\beta_{i,a,1};
b_{2,i,a}, A_i
\right)
\Bigg].
\end{aligned}
\label{eq:closed_form_transition_b0}
\end{equation}
For $b=1,\ldots,M-2$,

\begin{equation}
\begin{aligned}
[\mathbf{T}_{i}^{\mathrm{nl,pw}}]_{a,b}
&=
\frac{\xi_i}{A_i^{\xi_i}}
\Big[
\mathcal{F}_{\xi_i}
\left(
\alpha_{i,a,b+1}; h_{L,i}, b_{1,i,a}
\right)
\\
&\quad
-
\mathcal{F}_{\xi_i}
\left(
\alpha_{i,a,b}; h_{L,i}, b_{1,i,a}
\right)
\\
&\quad
+
\mathcal{C}_{\xi_i}
\left(
\beta_{i,a,b+1}; b_{2,i,a}, A_i
\right)
\\
&\quad
-
\mathcal{C}_{\xi_i}
\left(
\beta_{i,a,b}; b_{2,i,a}, A_i
\right)
\Big].
\end{aligned}
\label{eq:closed_form_transition_interior}
\end{equation}
For $b=M-1$,

\begin{equation}
\begin{aligned}
[\mathbf{T}_{i}^{\mathrm{nl,pw}}]_{a,M-1}
&=
\frac{\xi_i}{A_i^{\xi_i}}
\Big[
\mathcal{P}_{\xi_i}
\left(
h_{L,i}, b_{1,i,a}
\right)
\\
&\quad
-
\mathcal{F}_{\xi_i}
\left(
\alpha_{i,a,M-1};
h_{L,i}, b_{1,i,a}
\right)
\\
&\quad
+
\mathcal{P}_{\xi_i}
\left(
b_{2,i,a}, A_i
\right)
\\
&\quad
-
\mathcal{C}_{\xi_i}
\left(
\beta_{i,a,M-1};
b_{2,i,a}, A_i
\right)
\Big].
\end{aligned}
\label{eq:closed_form_transition_last}
\end{equation}
Equations~\eqref{eq:closed_form_transition_b0}--\eqref{eq:closed_form_transition_last}
approximate the exact non-outage transition contribution in
\eqref{eq:non_outage_transition_matrix_expanded} under the same signal--ASE and
constant-noise partition used for the closed-form SER in
\eqref{eq:signal_ase_integral_closed_form}. The corresponding closed-form
per-hop transition matrix is
\begin{equation}
\mathbf{T}_{i}^{\mathrm{pw}}
=
P_{\mathrm{out},i}\mathbf{T}_{i}^{\mathrm{out}}
+
\mathbf{T}_{i}^{\mathrm{nl,pw}} .
\label{eq:closed_form_per_hop_transition_matrix}
\end{equation}
If $D_{i,a}=0$, the signal--ASE-dominant interval is absent for row $a$, and
the corresponding entries are obtained by retaining only the constant-noise
terms over $[h_{L,i},A_i]$.

\subsection{End-to-End Transition Matrix and SER}

The end-to-end transition matrix over the selected $H$-hop relay chain is
\begin{equation}
\mathbf{T}_{\mathrm{tot}}
=
\mathbf{T}_1\mathbf{T}_2\cdots\mathbf{T}_H,
\label{eq:end_to_end_transition_matrix}
\end{equation}
where $\mathbf{T}_i$ is the per-hop transition matrix defined in
\eqref{eq:per_hop_transition_matrix}--\eqref{eq:transition_matrix_decomposition}.
Hence, the conditional end-to-end transition probability from source symbol $a$
to final detected symbol $b$ is
\begin{equation}
\Pr\{X_H=b \mid X_0=a\}
=
[\mathbf{T}_{\mathrm{tot}}]_{a,b},
\qquad
a,b\in\mathcal{S}.
\label{eq:end_to_end_transition_probability}
\end{equation}
The conditional end-to-end SER for source symbol $a$ is therefore
\begin{equation}
P_{s,\mathrm{e2e}}(a)
=
1-[\mathbf{T}_{\mathrm{tot}}]_{a,a},
\qquad
a\in\mathcal{S}.
\label{eq:conditional_end_to_end_ser}
\end{equation}
For equiprobable source symbols, the average end-to-end SER is

\begin{equation}
\overline{P}_{s,\mathrm{e2e}}
=
1
-
\frac{1}{M}
\sum_{a=0}^{M-1}
[\mathbf{T}_{\mathrm{tot}}]_{a,a}.
\label{eq:average_end_to_end_ser}
\end{equation}
Equivalently, using the trace operator,
\eqref{eq:average_end_to_end_ser} can be written compactly as
\begin{equation}
\overline{P}_{s,\mathrm{e2e}}
=
1
-
\frac{1}{M}
\operatorname{tr}
\left(
\mathbf{T}_{\mathrm{tot}}
\right).
\label{eq:trace_end_to_end_ser}
\end{equation}
If the two-region closed-form transition model is used, the corresponding
end-to-end transition matrix is
\begin{equation}
\mathbf{T}_{\mathrm{tot}}^{\mathrm{pw}}
=
\mathbf{T}_{1}^{\mathrm{pw}}
\mathbf{T}_{2}^{\mathrm{pw}}
\cdots
\mathbf{T}_{H}^{\mathrm{pw}},
\label{eq:closed_form_end_to_end_transition_matrix}
\end{equation}
where $\mathbf{T}_{i}^{\mathrm{pw}}$ is given in
\eqref{eq:closed_form_per_hop_transition_matrix}. The closed-form end-to-end
SER approximation is then
\begin{equation}
\overline{P}_{s,\mathrm{e2e}}^{\mathrm{pw}}
=
1
-
\frac{1}{M}
\operatorname{tr}
\left(
\mathbf{T}_{\mathrm{tot}}^{\mathrm{pw}}
\right).
\label{eq:closed_form_end_to_end_ser}
\end{equation}

Equations~\eqref{eq:end_to_end_transition_matrix}--\eqref{eq:closed_form_end_to_end_ser}
show that the multi-hop performance is determined by the product of per-hop
symbol-transition matrices. Thus, the proposed regenerative architecture
converts the end-to-end error analysis from waveform-level noise accumulation
into discrete-state error propagation across the relay chain.

\begin{figure}[!t]
    \centering
    \includegraphics[width=\columnwidth]{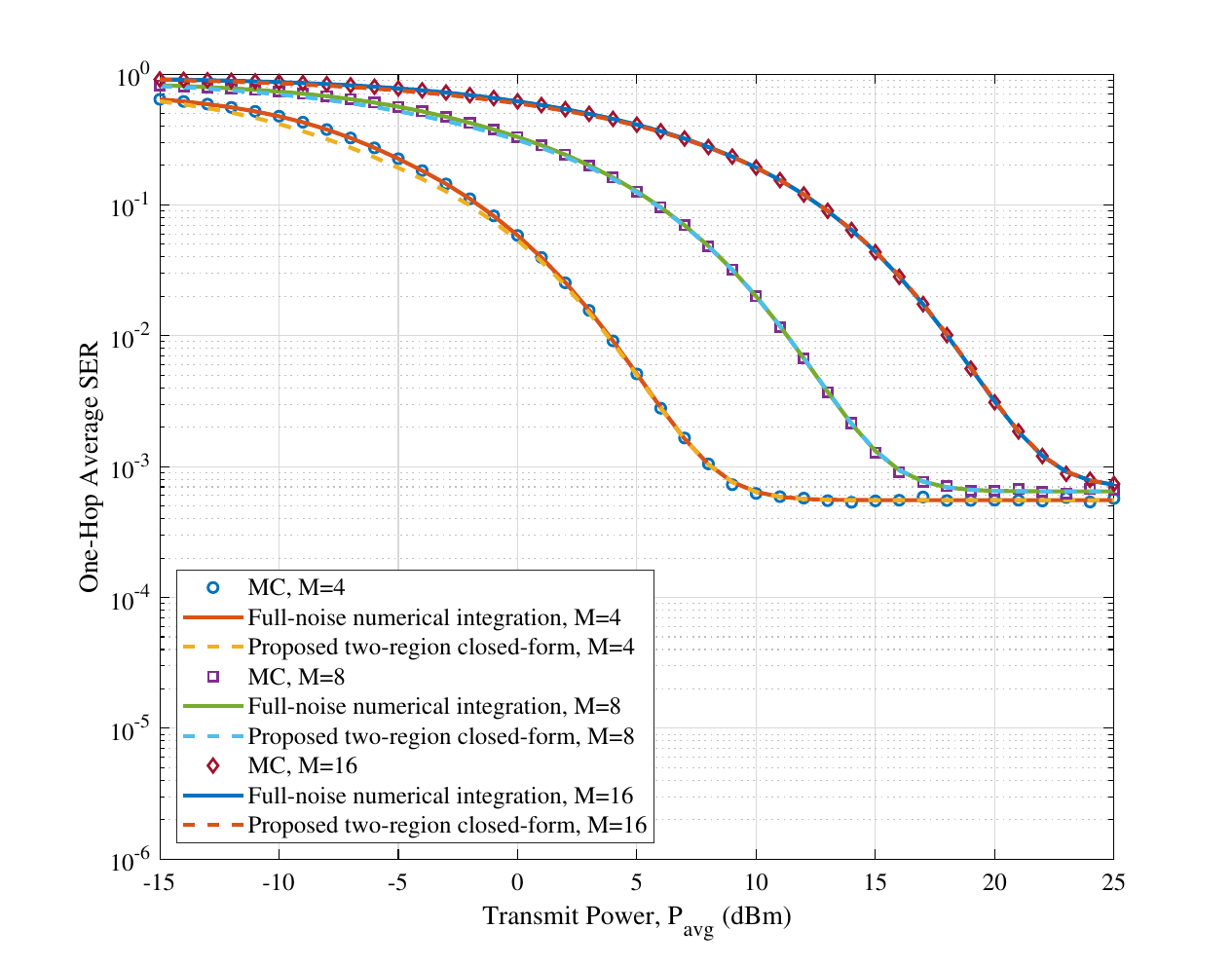}
\caption{One-hop average SER versus transmit power for $M = 4$, $8$, and $16$ PAM orders, comparing MC simulation, full-noise numerical integration, and the proposed two-region closed-form analysis under the one-hop validation setup with $L = 600~\mathrm{km}$ and $\omega_{0} = 0.0675~\mathrm{m}$.}
    \label{fig:fig2}
\end{figure}

\subsection{Identically Distributed Hops}

When all hops have identical channel statistics and hardware parameters, the
per-hop transition matrices become identical, i.e.,
\begin{equation}
\mathbf{T}_i
=
\mathbf{T},
\qquad
i=1,\ldots,H,
\label{eq:identical_per_hop_transition_matrices}
\end{equation}
where $\mathbf{T}$ denotes the common per-hop transition matrix. In this case,
the end-to-end transition matrix in \eqref{eq:end_to_end_transition_matrix}
reduces to
\begin{equation}
\mathbf{T}_{\mathrm{tot}}
=
\mathbf{T}^{H},
\label{eq:identical_hop_end_to_end_transition_matrix}
\end{equation}
where $H$ is the total number of hops defined in \eqref{eq:number_of_hops}.
Therefore, the average end-to-end SER becomes
\begin{equation}
\overline{P}_{s,\mathrm{e2e}}
=
1
-
\frac{1}{M}
\operatorname{tr}
\left(
\mathbf{T}^{H}
\right).
\label{eq:identical_hop_end_to_end_ser}
\end{equation}

Let $\{\lambda_{\ell}\}_{\ell=1}^{M}$ denote the eigenvalues of $\mathbf{T}$,
counted with algebraic multiplicity. Since
\begin{equation}
\operatorname{tr}
\left(
\mathbf{T}^{H}
\right)
=
\sum_{\ell=1}^{M}
\lambda_{\ell}^{H},
\label{eq:trace_eigenvalue_expansion}
\end{equation}
the end-to-end SER can be equivalently expressed as
\begin{equation}
\overline{P}_{s,\mathrm{e2e}}
=
1
-
\frac{1}{M}
\sum_{\ell=1}^{M}
\lambda_{\ell}^{H}.
\label{eq:eigenvalue_end_to_end_ser}
\end{equation}
Equation~\eqref{eq:eigenvalue_end_to_end_ser} shows that, for statistically
identical hops, the multi-hop error behavior is governed by the eigenstructure
of the per-hop transition matrix. As $H$ increases, the dominant eigenmodes
determine how the regenerated symbol distribution evolves along the relay
chain. If the two-region closed-form transition model is used, the same
expressions hold by replacing $\mathbf{T}$ with $\mathbf{T}^{\mathrm{pw}}$.

\begin{figure}[!t]
    \centering
    \includegraphics[width=\columnwidth]{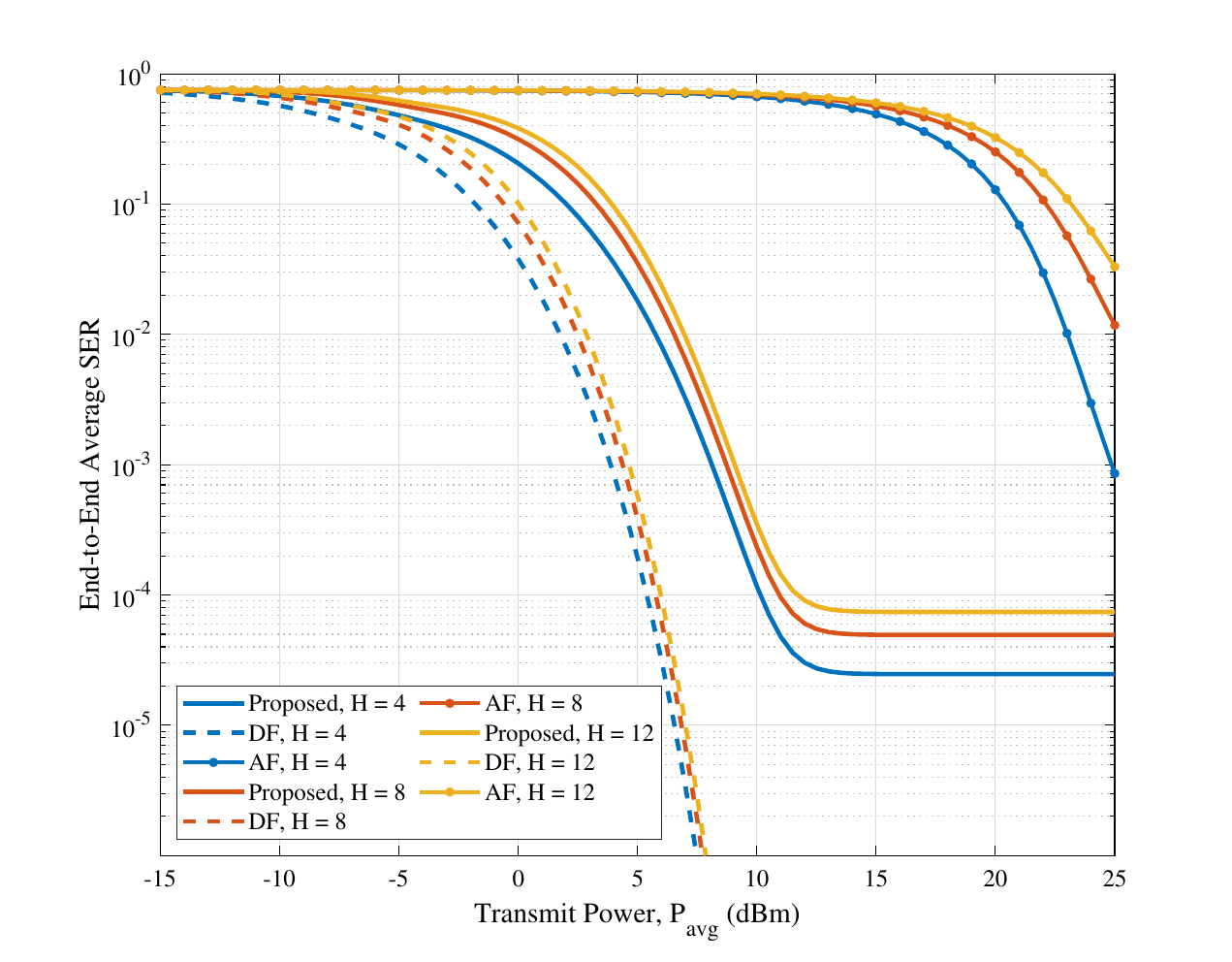}
\caption{End-to-end average SER versus transmit power for the proposed all-optical regenerative relay compared with DF and AF relaying for $M = 4$ and $H = 4, 8,$ and $12$, under $L = 500~\mathrm{km}$ and $\omega_{0} = 0.05~\mathrm{m}$.}
    \label{fig:fig3}
\end{figure}

\begin{table}[!t]
\centering
\caption{Values of Simulation Parameters}
\label{tab:tab1}
\renewcommand{\arraystretch}{1.15}
\begin{tabular}{|c c|c c|}
\hline
\textbf{Parameter} & \textbf{Value} & \textbf{Parameter} & \textbf{Value} \\
\hline
$\lambda$ & $1550~\mathrm{nm}$ 
& $L$ & $600~\mathrm{km}$ \\

$w_0$ & $0.0675~\mathrm{m}$ 
& $w_L$ & $1$--$10~\mathrm{m}$ \\

$a$ & $0.05~\mathrm{m}$ 
& $\sigma_{\theta}$ & $1$--$4~\mu\mathrm{rad}$ \\

$M$ & $4,~8,~16$ 
& $P_{\max}$ & $-15$ to $25~\mathrm{dBm}$ \\

$G_{\mathrm{TX}}$ & $10$ 
& $G_{\max}$ & $10^3$ \\

$B_o$ & $50~\mathrm{GHz}$ 
& $B_e$ & $25~\mathrm{GHz}$ \\
\hline
\end{tabular}
\end{table}

\section{Simulation Results and Discussion}

This section presents simulation results for the proposed all-optical
regenerative inter-satellite relay. Unless otherwise stated, the parameters in
Table~\ref{tab:tab1} are used, including the link length, aperture/beam geometry, pointing
jitter, finite EDFA gain range, filtering bandwidths, ASE-related noise,
background noise, and decision-stage noise. The transmit-power range and
$w_0=0.0675~\mathrm{m}$ are selected to match practical
laser-communication-terminal assumptions; in particular, $w_0$ corresponds to
an aperture-consistent Gaussian-beam parameter for a $135$-mm-class optical
aperture, as in EDRS/TESAT-class terminals \cite{tesat2024mpb100gbps}. The closed-form SER is first
validated against MC simulation and full-noise numerical integration,
and is then used to study transmit power, modulation order, pointing jitter,
beamwidth, hop count, EDFA limits, and threshold scaling.

\begin{figure*}[!t]
    \centering

    \subfloat[]{%
        \includegraphics[width=0.48\textwidth]{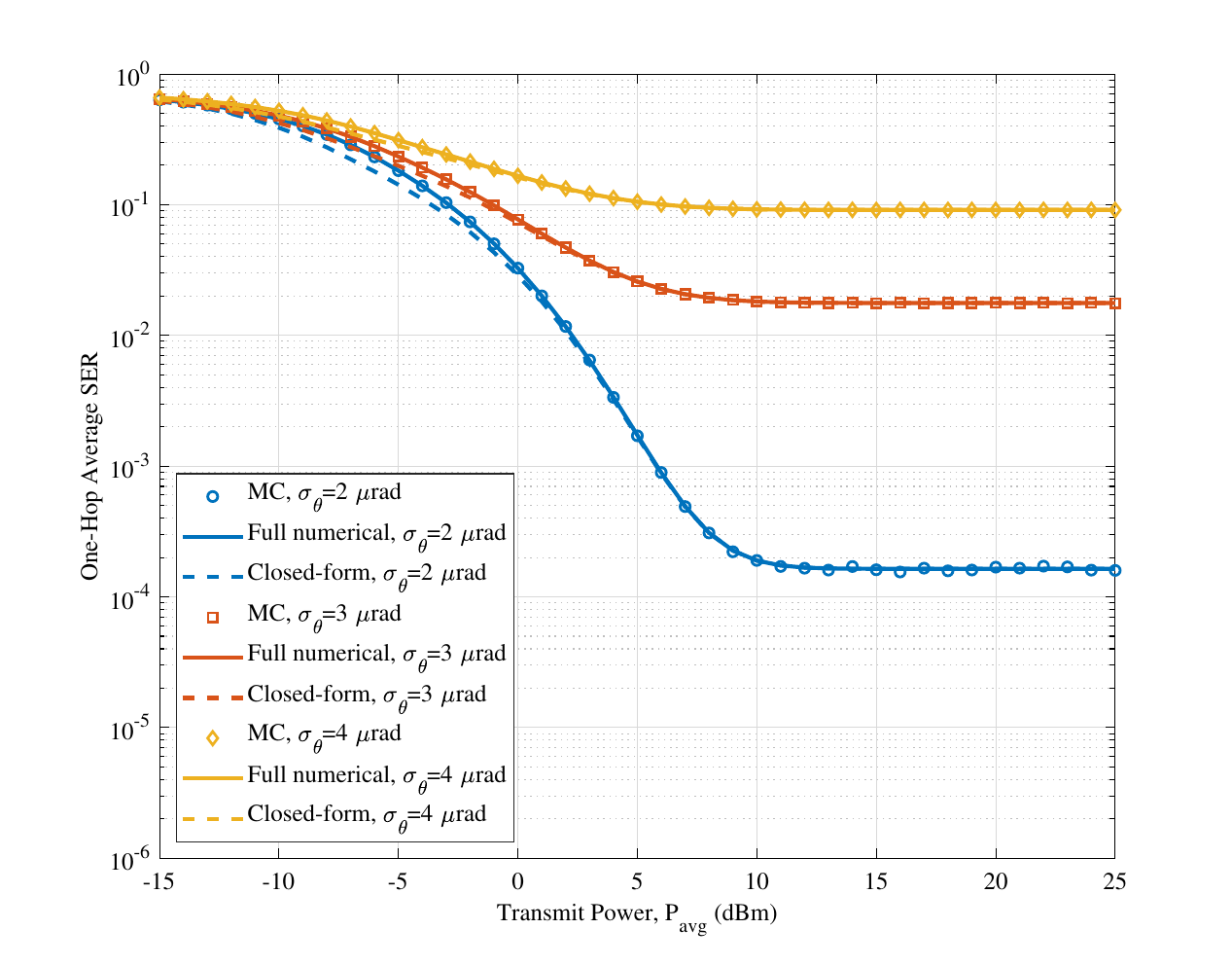}
        \label{fig:fig4a}
    }
    \hfill
    \subfloat[]{%
        \includegraphics[width=0.48\textwidth]{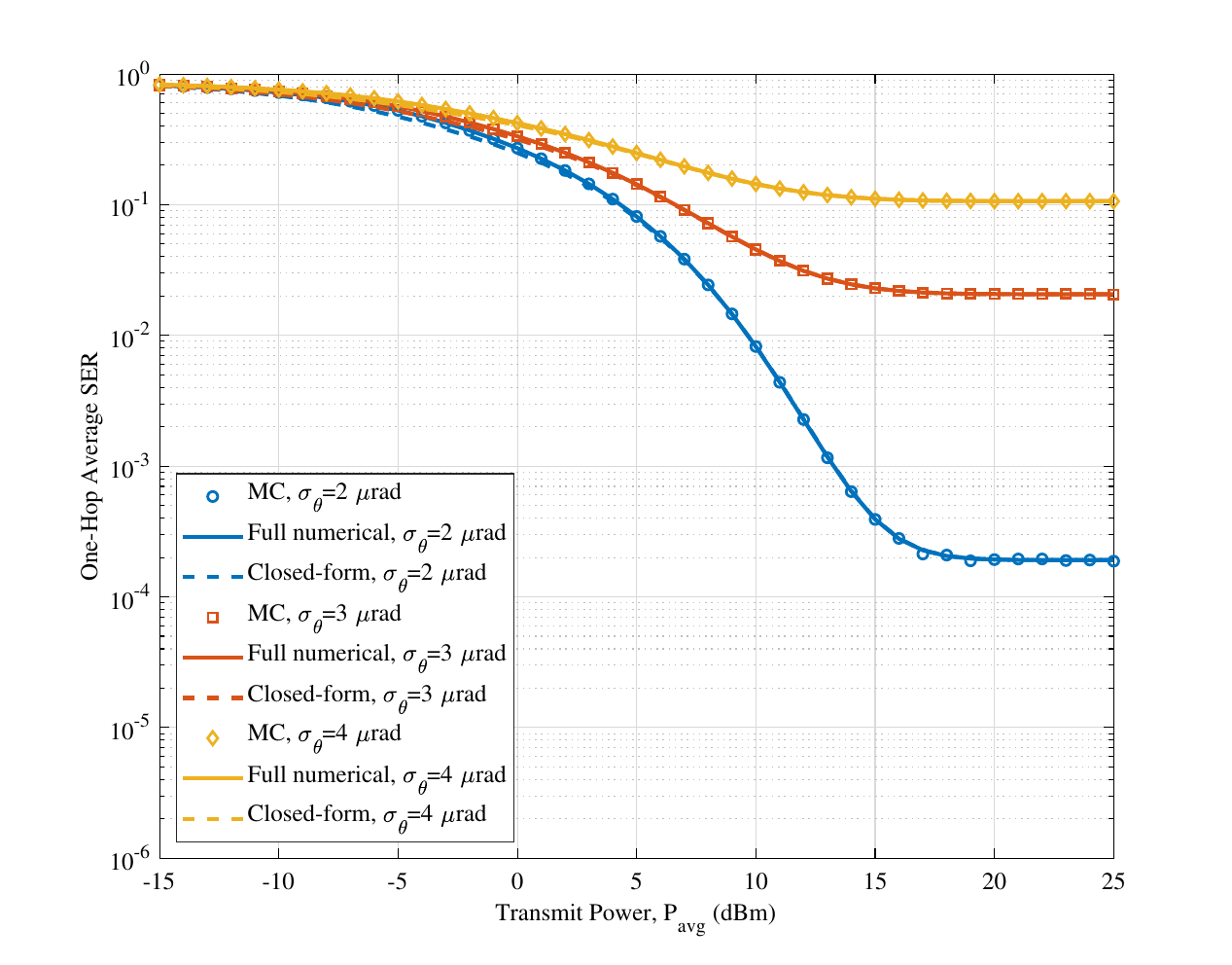}
        \label{fig:fig4b}
    }

    \caption{One-hop average SER versus transmit power for $\sigma_{\theta}=2$, $3$, and $4~\mu\mathrm{rad}$, comparing MC, full-noise numerical integration, and closed-form analysis for (a) $M=4$ and (b) $M=8$.}
    \label{fig:fig4}
\end{figure*}

\begin{figure}[!t]
    \centering
    \includegraphics[width=\columnwidth]{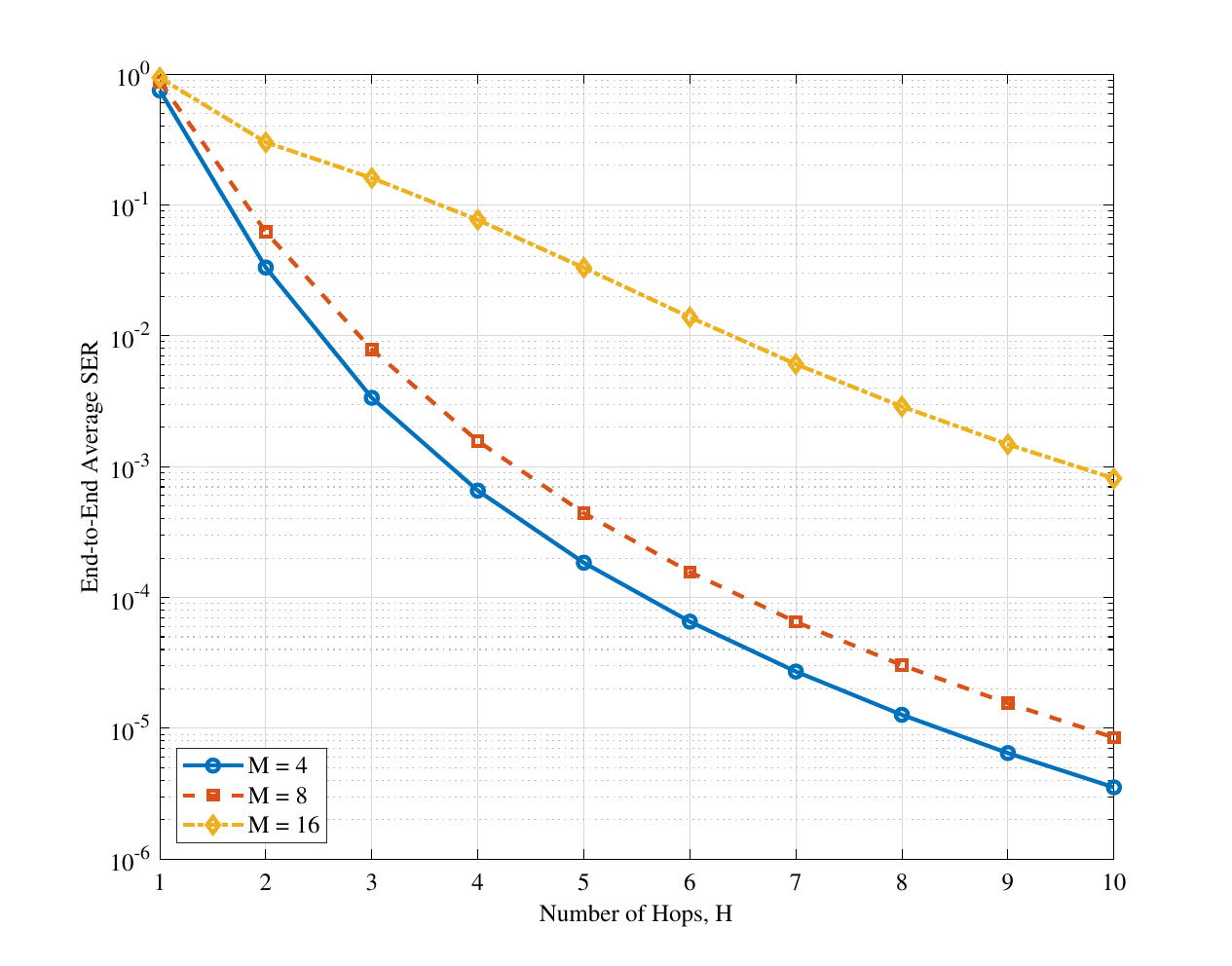}
    \caption{ End-to-end average SER versus hop count for  $M=4$, $8$, and $16$.}
    \label{fig:fig5}
\end{figure}

Fig.~2 validates the one-hop SER analysis for $M=4$, $8$, and $16$: MC
simulation, full-noise numerical integration, and the proposed two-region closed
form nearly overlap across the power range. As $M$ increases, reduced M-PAM
spacing shifts the waterfall to higher $P_{\mathrm{avg}}$. The high-power floor
appears at about $10$, $17$, and $24~\mathrm{dBm}$ for $M=4$, $8$, and $16$,
respectively, implying roughly a $7~\mathrm{dB}$ power penalty per
modulation-order step. The nearly common floor, around $7\times 10^{-4}$,
confirms that deep pointing fades and gain-limited outage set the ultimate
one-hop reliability, so further transmit-power increase gives little additional
improvement.

Fig.~3 compares the end-to-end SER of the proposed regenerative relay with DF
and AF for $H=4$, $8$, and $12$. The proposed relay shows a waterfall response
followed by low floors of about $2\times 10^{-5}$, $5\times 10^{-5}$, and
$8\times 10^{-5}$, respectively, as more hops increase regenerated
decision-error propagation. For $H=4$, it reaches the $10^{-4}$ region at about
$11~\mathrm{dBm}$, with an approximately $5~\mathrm{dB}$ penalty relative to DF.
This gap is expected because DF performs full O/E/O symbol recovery and clean
retransmission. In contrast, AF remains noise-limited because it forwards and
re-amplifies the analog noise together with the signal; at $25~\mathrm{dBm}$,
its SER is about $10^{-3}$, $10^{-2}$, and $3\times 10^{-2}$ for $H=4$, $8$,
and $12$. Thus, the proposed relay retains all-optical operation while strongly
suppressing AF-like noise accumulation.

Fig.~4 evaluates the effect of pointing jitter on the one-hop SER for $M=4$ and
$M=8$. The MC, full-noise numerical integration, and closed-form curves
remain closely aligned for all $\sigma_\theta$, confirming the accuracy of the
analytical model under different pointing-error severities. Increasing
$\sigma_\theta$ mainly raises the high-power floor: for $M=4$, the floor
increases from about $1.5\times 10^{-4}$ at $2~\mu\mathrm{rad}$ to
$1.8\times 10^{-2}$ and $9\times 10^{-2}$ at $3$ and $4~\mu\mathrm{rad}$,
respectively. For $M=8$, the waterfall shifts right by about $7~\mathrm{dB}$,
and the floors are slightly higher. Hence, beyond the waterfall region,
reliability is limited by gain-limited outage caused by deep pointing fades
rather than transmit power.

Fig.~5 shows the end-to-end SER when a fixed total distance is partitioned into
more regenerative hops. Increasing $H$ shortens each hop, improving aperture
coupling and reducing pointing-induced fading; this gain dominates the
additional regenerated decision stages, so the SER decreases for all modulation
orders. The improvement is rapid for lower-order PAM: by $H=4$, the SER drops to
about $7\times 10^{-4}$ for $M=4$ and $1.5\times 10^{-3}$ for $M=8$, whereas
$M=16$ remains near $8\times 10^{-2}$ because of its narrower decision margins.
At $H=10$, the SER reaches about $4\times 10^{-6}$, $1\times 10^{-5}$, and
$9\times 10^{-4}$ for $M=4$, $8$, and $16$, respectively. Thus, hop
partitioning mitigates long-distance loss and pointing impairment, with stronger
reliability gains for lower-order PAM.

\begin{figure*}[!t]
    \centering

    \subfloat[]{%
        \includegraphics[width=0.48\textwidth]{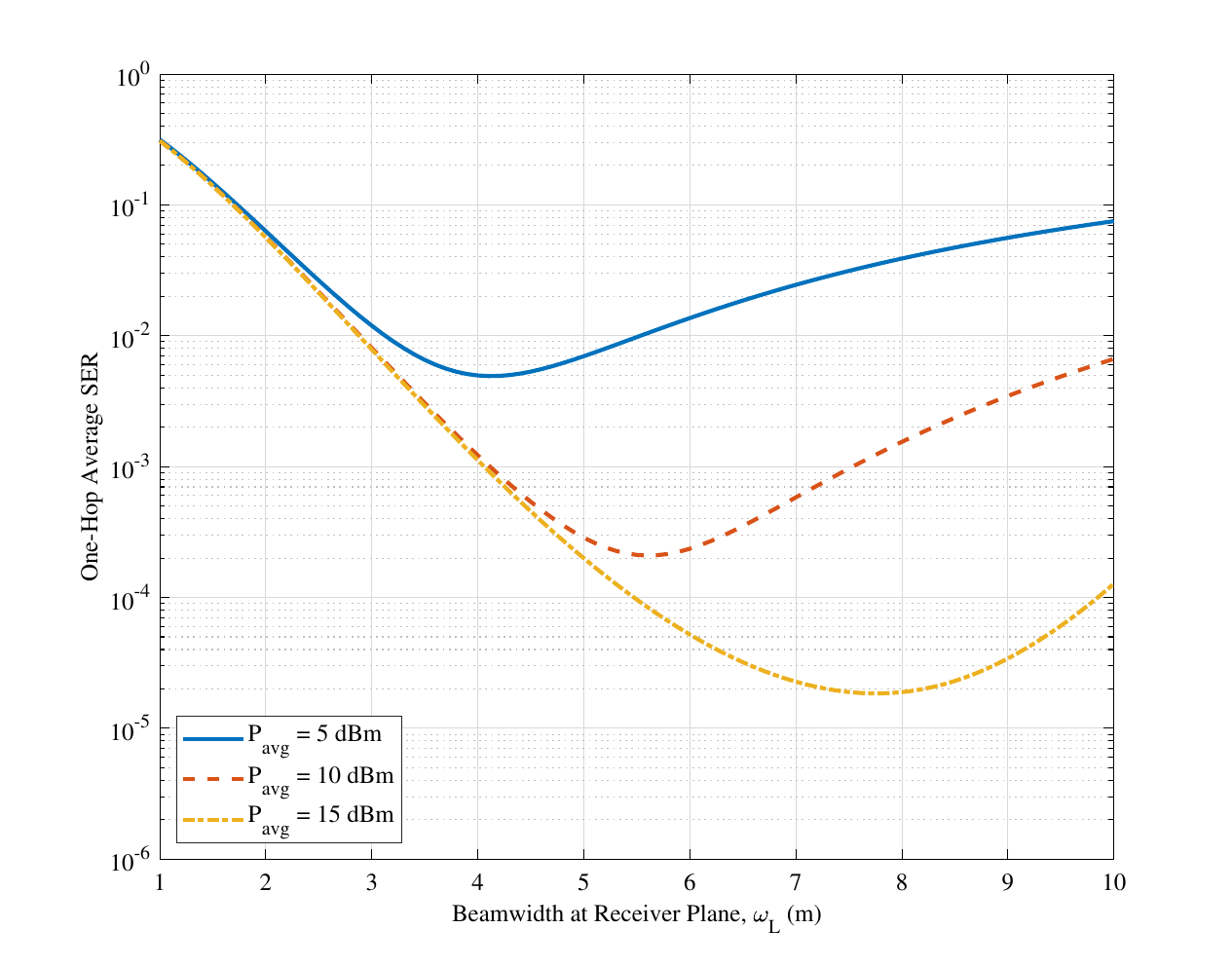}
        \label{fig:fig6a}
    }
    \hfill
    \subfloat[]{%
        \includegraphics[width=0.48\textwidth]{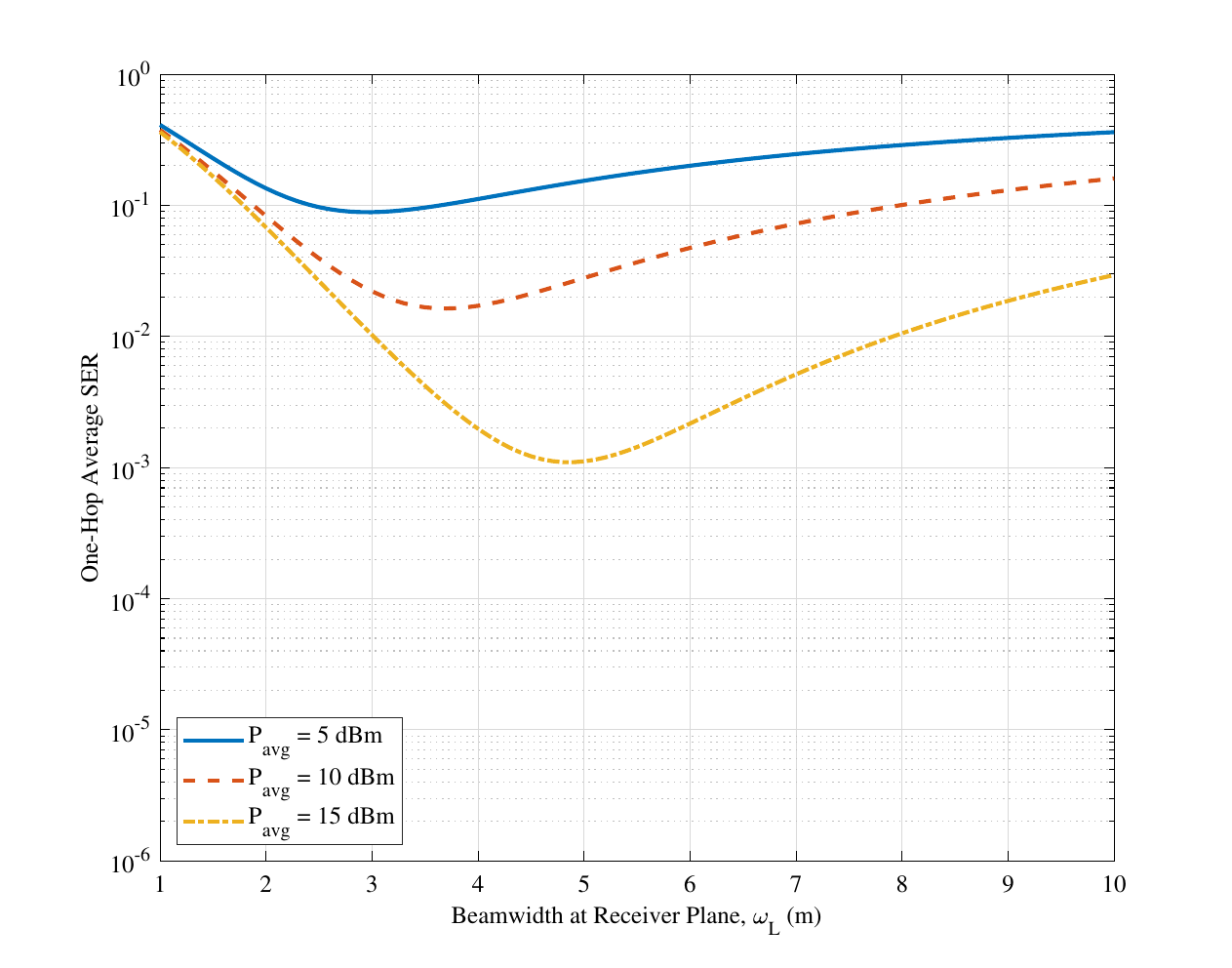}
        \label{fig:fig6b}
    }

    \caption{One-hop average SER versus beamwidth at the receiver plane for $P_{\max}=5$, $10$, and $15~\mathrm{dBm}$, showing the beamwidth-dependent performance for (a) $M=4$ and (b) $M=8$.}
    \label{fig:fig6}
\end{figure*}

\begin{figure*}[!t]
    \centering

    \subfloat[]{%
        \includegraphics[width=0.31\textwidth]{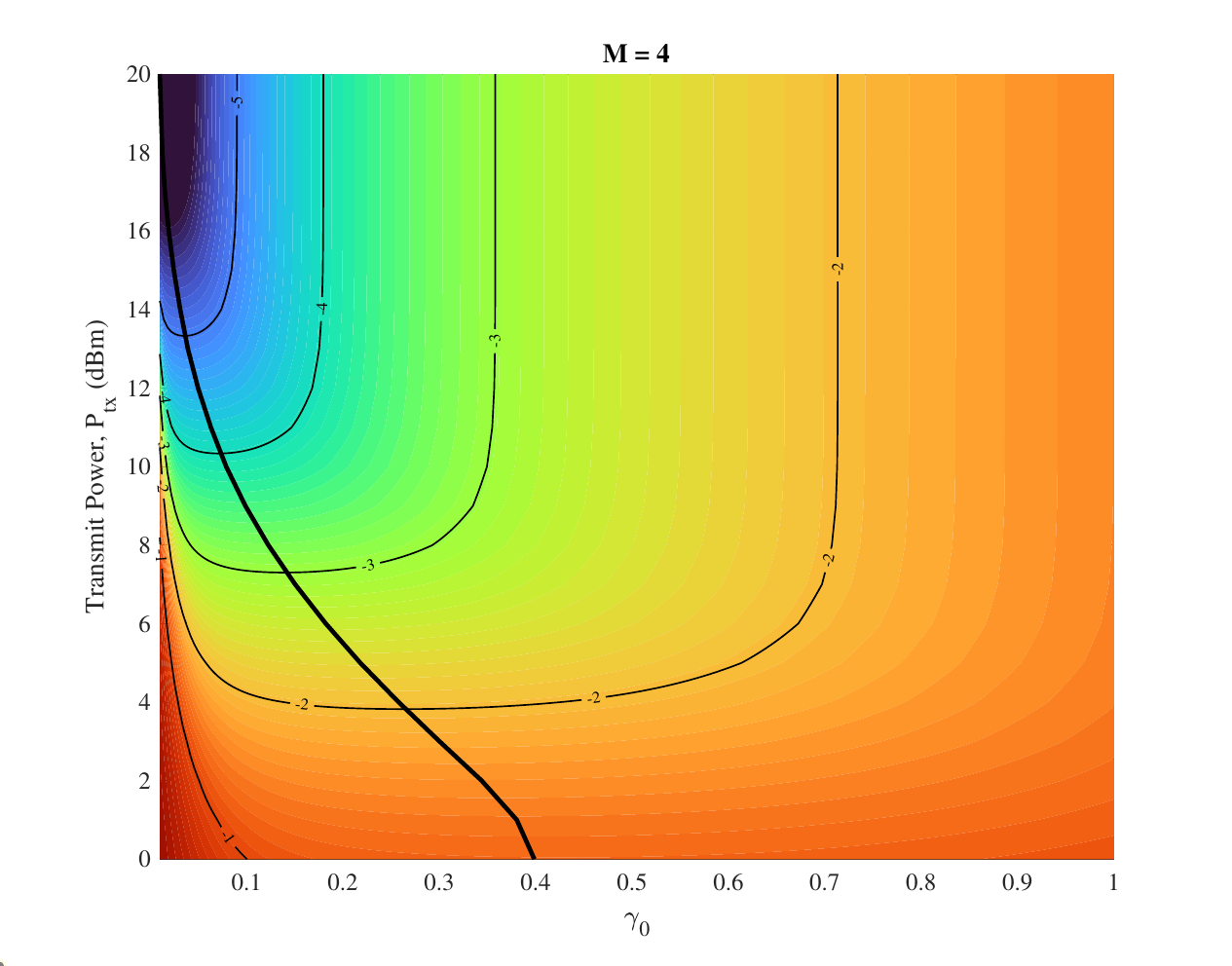}
        \label{fig:fig7a}
    }
    \hfill
    \subfloat[]{%
        \includegraphics[width=0.31\textwidth]{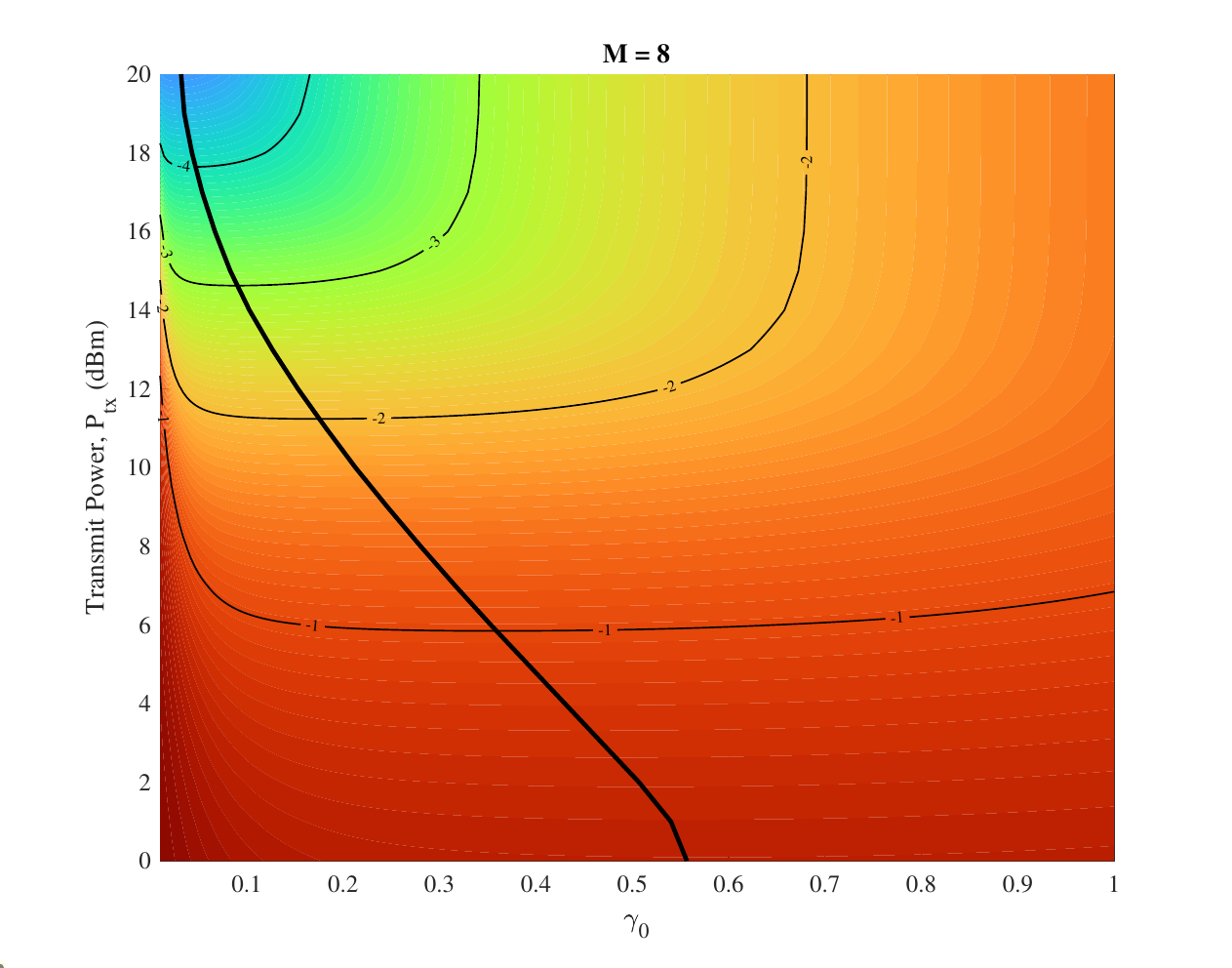}
        \label{fig:fig7b}
    }
     \hfill
    \subfloat[]{%
        \includegraphics[width=0.31\textwidth]{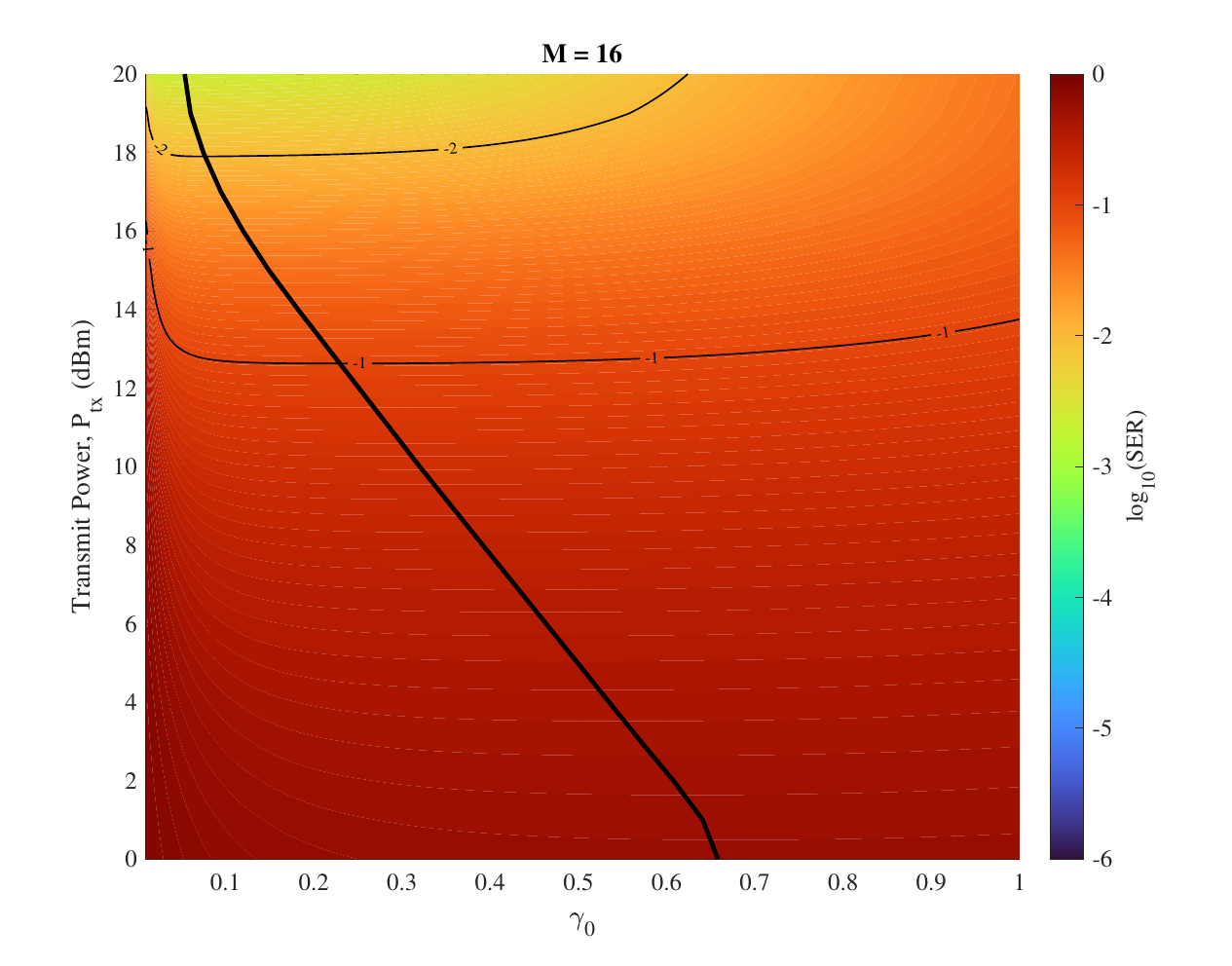}
        \label{fig:fig7c}
    }
    
    \caption{One-hop SER landscape versus threshold-scaling factor $\gamma_{0}$ and transmit power $P_{\mathrm{tx}}$ for (a) $M=4$, (b) $M=8$, and (c) $M=16$.}
    \label{fig:fig7}
\end{figure*}

Fig.~6 shows that beamwidth has a non-monotonic impact on the one-hop SER
because it trades aperture coupling against pointing robustness. Narrow beams
provide stronger collected intensity but are highly sensitive to pointing
displacement, whereas overly wide beams reduce the collected optical power;
hence, each curve has an optimum beamwidth. For $M=4$, the minimum SER improves
from about $5\times 10^{-3}$ at $P_{\max}=5~\mathrm{dBm}$ to
$2\times 10^{-4}$ and $2\times 10^{-5}$ at $P_{\max}=10$ and $15~\mathrm{dBm}$,
respectively, while the optimum $w_L$ shifts to larger values as power
increases. For $M=8$, the same U-shaped trend appears, but the curves remain
higher and the optimum beamwidth is smaller because the reduced M-PAM spacing
makes the OHL decisions more sensitive to beam-expansion loss. Thus, transmit
power improves the optimum SER, but proper beamwidth selection is still required
to avoid both pointing-dominated and collection-limited operation.

Fig.~7 shows the per-hop SER landscape versus the gain target $\gamma_0$ and
transmit power for $M=4$, $8$, and $16$. The reliable region shrinks as $M$
increases because higher-order PAM reduces the OHL decision margin and increases
sensitivity to residual noise and gain mismatch. For $M=4$, SER values near
$10^{-4}$ are obtained around $P_{\mathrm{tx}}=13~\mathrm{dBm}$ when $\gamma_0$
is small, whereas increasing $\gamma_0$ beyond about $0.35$ moves the SER back
toward the $10^{-3}$--$10^{-2}$ region. For $M=8$, comparable low-SER operation
requires about $17~\mathrm{dBm}$ and remains confined to small $\gamma_0$. For
$M=16$, the landscape is mostly limited to the $10^{-2}$--$10^{-1}$ range even
at high power. The thick black boundary marks the gain-limited condition,
showing that overly large $\gamma_0$ increases EDFA gain demand and can make the
SER outage-dominated.

Fig.~8 visualizes the end-to-end transition matrix for $M=8$ after $H=3$
regenerative hops. The probability mass is concentrated around the main
diagonal, showing that most symbols are preserved through the relay chain rather
than being mapped to distant levels. Lower symbols are detected more reliably,
while middle and upper symbols show broader adjacent-symbol spreading; for
example, the diagonal probability decreases to about $0.775$ for $a=5$ and
$0.710$ for $a=6$, with dominant errors going to neighboring levels. The
nonzero $b=0$ column, around $2.3\times 10^{-4}$ for several source symbols,
reflects rare gain-limited deep fades that collapse the detected level. Thus,
multi-hop errors are structured, mainly consisting of neighboring-level
confusion plus a small outage-induced collapse component.

\begin{figure}[!t]
    \centering
    \includegraphics[width=\columnwidth]{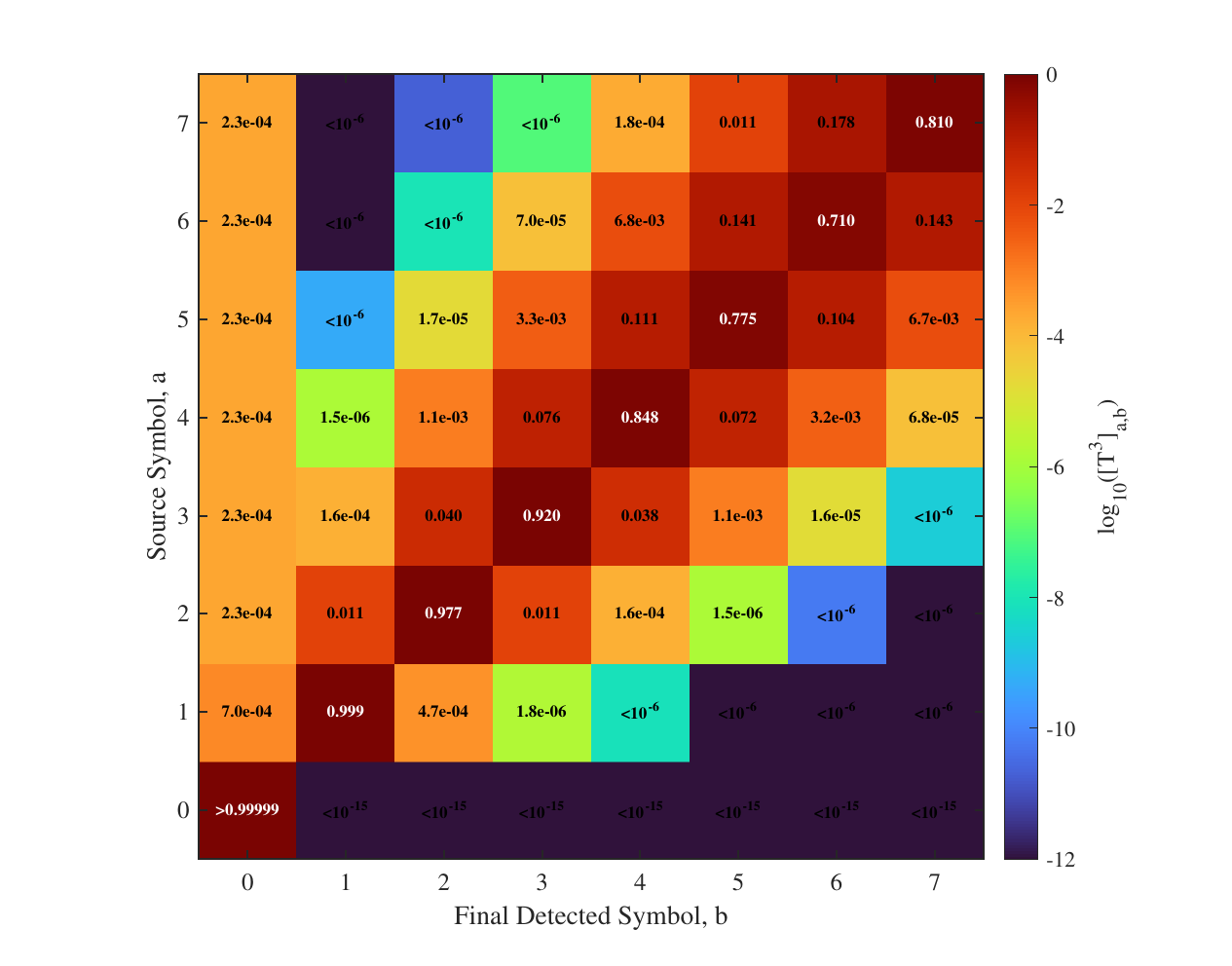}
    \caption{End-to-end transition-probability matrix for $M=8$ after $H=3$ regenerative hops, showing the probability of detecting symbol $b$ given source symbol $a$.}
    \label{fig:fig8}
\end{figure}

\section{Conclusion}
This paper proposed an all-optical regenerative relaying framework for M-PAM multi-hop inter-satellite optical links under pointing errors. The relay uses a parallel OHL bank for symbol-level optical regeneration, avoiding AF noise accumulation while eliminating the O/E/O conversion, buffering, and electrical processing required by DF. A variable-gain EDFA mechanism was developed to align received M-PAM levels with fixed optical thresholds, and the threshold-stable region and gain-limited outage condition were characterized. Based on a comprehensive optical-noise model, closed-form per-hop SER expressions were derived and extended to end-to-end performance through a Markov transition-matrix framework. MC simulations and full-noise numerical integration showed near-exact agreement with the analysis. The results showed that higher-order M-PAM improves spectral efficiency but reduces decision margins and increases sensitivity to pointing errors and noise. Transmit power improves SER only up to an outage-limited floor caused by deep pointing fades and finite EDFA gain. The analysis also highlighted the roles of beamwidth, pointing accuracy, threshold scaling, EDFA gain range, and hop count. Overall, the framework provides a benchmark for low-latency, spectrally efficient, and pointing-aware all-optical regenerative relaying in future LEO inter-satellite optical networks.

\bibliographystyle{IEEEtran}
\bibliography{references}

@article{balakrishnan2025toward,
  author  = {S. Balakrishnan and G. Thangavel and M. Manikandan and S. Vinodhkumar and M. Rajesh and M. Shabaz},
  title   = {Toward Standardized Energy-Aware Adaptive Routing Protocols for Ultra-Dense 6G Networks With Integrated Satellite-Terrestrial Architectures},
  journal = {IEEE Communications Standards Magazine},
  year    = {2025},
  pages   = {1--11},
  doi     = {10.1109/MCOMSTD.2025.3605636}
}

@article{mouhammad2026optimal,
  author  = {C. Mouhammad and M. S. Bashir and M. S. Alouini},
  title   = {Optimal Satellite Orbits for Delay-Tolerant Free-Space Optical Networks},
  journal = {IEEE Transactions on Aerospace and Electronic Systems},
  year    = {2026},
  pages   = {1--19},
  doi     = {10.1109/TAES.2026.3692759}
}

@article{ghanbari2026narrowbeams,
  author  = {Ghanbari, M. and Dabiri, M. T. and Badarneh, O. S. and Hasna, M. and Al-Badarneh, Y. H. and Alshawaqfeh, M. K. and Qaraqe, K.},
  title   = {When Future Communications Shift Toward Narrow Beams: A Forward-Looking Survey on Pointing Errors and Alignment Limits},
  journal = {IEEE Open Journal of the Communications Society},
  volume  = {7},
  pages   = {4959--5006},
  year    = {2026},
  doi     = {10.1109/OJCOMS.2026.3690915}
}

@article{dabiri2018channel,
  author  = {Dabiri, M. T. and Sadough, S. M. S. and Khalighi, M. A.},
  title   = {Channel Modeling and Parameter Optimization for Hovering UAV-Based Free-Space Optical Links},
  journal = {IEEE Journal on Selected Areas in Communications},
  volume  = {36},
  number  = {9},
  pages   = {2104--2113},
  year    = {2018},
  doi     = {10.1109/JSAC.2018.2864416}
}

@article{liu2020relay,
  author  = {Liu, W. and Ding, J. and Zheng, J. and Chen, X.},
  title   = {Relay-Assisted Technology in Optical Wireless Communications: A Survey},
  journal = {IEEE Access},
  volume  = {8},
  pages   = {194384--194409},
  year    = {2020},
  doi     = {10.1109/ACCESS.2020.3032930}
}

@article{dabiri2018allopticalaf,
  author  = {Dabiri, M. T. and Sadough, S. M. S.},
  title   = {Performance Analysis of All-Optical Amplify and Forward Relaying Over Log-Normal FSO Channels},
  journal = {Journal of Optical Communications and Networking},
  volume  = {10},
  number  = {2},
  pages   = {79--89},
  month   = feb,
  year    = {2018},
  doi     = {10.1364/JOCN.10.000079}
}

@article{erdogan2022secrecy,
  author  = {Erdogan, E. and Altunbas, I. and Kurt, G. K. and Yanikomeroglu, H.},
  title   = {The Secrecy Comparison of RF and FSO Eavesdropping Attacks in Mixed RF-FSO Relay Networks},
  journal = {IEEE Photonics Journal},
  volume  = {14},
  number  = {1},
  pages   = {1--8},
  month   = feb,
  year    = {2022},
  doi     = {10.1109/JPHOT.2021.3127397}
}

@article{choudhary2024isowc,
  author  = {Choudhary, A. and Agrawal, N. K.},
  title   = {Inter-Satellite Optical Wireless Communication ({IsOWC}) Systems Challenges and Applications: A Comprehensive Review},
  journal = {Journal of Optical Communications},
  volume  = {45},
  number  = {4},
  pages   = {925--935},
  year    = {2024},
  doi     = {10.1515/joc-2022-0075}
}

@article{li2026incrementalhybrid,
  author  = {Li, J. and others},
  title   = {Performance Analysis and Optimization of Relay-Assisted Free-Space Optical Communications Based on Incremental Hybrid Decode-Amplify-Forward Scheme Over Atmospheric Turbulence},
  journal = {Journal of Lightwave Technology},
  volume  = {44},
  number  = {8},
  pages   = {2942--2950},
  month   = apr,
  year    = {2026},
  doi     = {10.1109/JLT.2026.3662416}
}

@article{yahia2022haps,
  author  = {Yahia, O. B. and Erdogan, E. and Kurt, G. K. and Altunbas, I. and Yanikomeroglu, H.},
  title   = {HAPS Selection for Hybrid RF/FSO Satellite Networks},
  journal = {IEEE Transactions on Aerospace and Electronic Systems},
  volume  = {58},
  number  = {4},
  pages   = {2855--2867},
  month   = aug,
  year    = {2022},
  doi     = {10.1109/TAES.2022.3142116}
}

@article{dabiri2021uavaf,
  author  = {Dabiri, M. T. and Khankalantary, S. and Piran, M. J. and Ansari, I. S. and Uysal, M. and Saad, W. and Hong, C. S.},
  title   = {UAV-Assisted Free Space Optical Communication System With Amplify-and-Forward Relaying},
  journal = {IEEE Transactions on Vehicular Technology},
  volume  = {70},
  number  = {9},
  pages   = {8926--8936},
  month   = sep,
  year    = {2021},
  doi     = {10.1109/TVT.2021.3098389}
}

@article{cai2019fewmodeedfa,
  author  = {Cai, S. and Zhang, Z. and Chen, X.},
  title   = {Turbulence-Resistant All Optical Relaying Based on Few-Mode {EDFA} in Free-Space Optical Systems},
  journal = {Journal of Lightwave Technology},
  volume  = {37},
  number  = {9},
  pages   = {2042--2049},
  month   = may,
  year    = {2019},
  doi     = {10.1109/JLT.2019.2897428}
}

@article{vu2018allopticaltwoway,
  author  = {Vu, M. Q. and Nguyen, N. T. T. and Pham, H. T. T. and Dang, N. T.},
  title   = {All-Optical Two-Way Relaying Free-Space Optical Communications for HAP-Based Broadband Backhaul Networks},
  journal = {Optics Communications},
  volume  = {410},
  pages   = {277--286},
  month   = mar,
  year    = {2018},
  doi     = {10.1016/j.optcom.2017.10.025}
}

@article{nor2017experimental,
  author  = {Nor, N. A. M. and Ghassemlooy, Z. and Bohata, J. and Saxena, P. and Komanec, M. and Zvanovec, S. and Bhatnagar, M. R. and Khalighi, M.-A.},
  title   = {Experimental Investigation of All-Optical Relay-Assisted 10 Gb/s FSO Link Over the Atmospheric Turbulence Channel},
  journal = {Journal of Lightwave Technology},
  volume  = {35},
  number  = {1},
  pages   = {45--53},
  month   = jan,
  year    = {2017},
  doi     = {10.1109/JLT.2016.2629081}
}

@article{trinh2015ohl,
  author  = {Trinh, P. V. and Dang, N. T. and Pham, A. T.},
  title   = {All-Optical Relaying FSO Systems Using EDFA Combined With Optical Hard-Limiter Over Atmospheric Turbulence Channels},
  journal = {Journal of Lightwave Technology},
  volume  = {33},
  number  = {19},
  pages   = {4132--4144},
  month   = oct,
  year    = {2015},
  doi     = {10.1109/JLT.2015.2466432}
}

@inproceedings{vu2016twowaync,
  author    = {Vu, M. Q. and Pham, H. T. T. and Pham, T. A. and Dang, N. T.},
  title     = {All-Optical Two-Way Relaying Dual-Hop {FSO} Systems Using Network Coding Over Atmospheric Turbulence Channel},
  booktitle = {2016 International Conference on Advanced Technologies for Communications (ATC)},
  address   = {Hanoi, Vietnam},
  pages     = {350--355},
  year      = {2016},
  doi       = {10.1109/ATC.2016.7764804}
}

@article{dabiri2025interstellarohl,
  author  = {Dabiri, M. T. and Hasna, M. and Althunibat, S. and Qaraqe, K.},
  title   = {All-Optical Inter-Satellite Relays With Intelligent Beam Control: Harnessing Liquid Lenses and Optical Hard Limiters},
  journal = {IEEE Transactions on Communications},
  volume  = {73},
  number  = {12},
  pages   = {14739--14752},
  month   = dec,
  year    = {2025},
  doi     = {10.1109/TCOMM.2025.3606646}
}

@article{farid2007outage,
  author  = {Farid, A. A. and Hranilovic, S.},
  title   = {Outage Capacity Optimization for Free-Space Optical Links With Pointing Errors},
  journal = {Journal of Lightwave Technology},
  volume  = {25},
  number  = {7},
  pages   = {1702--1710},
  month   = jul,
  year    = {2007},
  doi     = {10.1109/JLT.2007.899174}
}

@book{agrawal2010fiber,
  author    = {Agrawal, G. P.},
  title     = {Fiber-Optic Communication Systems},
  edition   = {4th},
  address   = {Hoboken, NJ, USA},
  publisher = {Wiley},
  year      = {2010}
}

@article{kahn2004spectral,
  author  = {Kahn, J. M. and Ho, K.-P.},
  title   = {Spectral Efficiency Limits and Modulation/Detection Techniques for {DWDM} Systems},
  journal = {IEEE Journal of Selected Topics in Quantum Electronics},
  volume  = {10},
  number  = {2},
  pages   = {259--272},
  month   = mar # {-} # apr,
  year    = {2004},
  doi     = {10.1109/JSTQE.2004.826575}
}

@book{proakis2008digital,
  author    = {Proakis, J. G. and Salehi, M.},
  title     = {Digital Communications},
  edition   = {5th},
  address   = {New York, NY, USA},
  publisher = {McGraw-Hill},
  year      = {2008}
}

@misc{tesat2024mpb100gbps,
  author       = {{TESAT}},
  title        = {{TESAT} and {MPB} Communications Successfully Demonstrate 100 {Gbps} Transmission Capability},
  howpublished = {Online},
  month        = apr,
  year         = {2024},
  note         = {Available: TESAT News}
}

\end{document}